\documentclass[letterpaper]{jpconf}
\usepackage{graphicx}
\usepackage{epsfig}
\def\lsim{\raise0.3ex\hbox{$<$\kern-0.75em\raise-1.1ex\hbox{$\sim$}}}
\def\gsim{\raise0.3ex\hbox{$>$\kern-0.75em\raise-1.1ex\hbox{$\sim$}}}

\def\mean#1{\left<#1\right>}
\def\Journal#1#2#3#4#5{ #5 {#1} {\it  {#2}} {\bf #3} #4}

\def\NPA{{Nucl. Phys. A}}
\def\NPB{{Nucl. Phys. B}}
\def\PLB{{Phys. Lett. B}}

\def\PRL{Phys. Rev. Lett.\ }
\def\ZPC{{Z. Phys. C}}
\def\ARNPS{{Ann. Rev. Nucl. Part. Sci.\ }}

\def\RPP{Rep. Prog. Phys.\ }
\begin{document}
\title{A fundamental test of the Higgs Yukawa coupling at RHIC in A+A collisions}
\author{M.~J.~Tannenbaum}
\address{Physics Department, 510c,
Brookhaven National Laboratory,
Upton, NY 11973-5000, USA}
\ead{mjt@bnl.gov}
\begin{abstract}
Searches for the intermediate boson, $W^{\pm}$, the heavy quantum of the Weak 
Interaction, via its semi-leptonic decay, $W\rightarrow e +\nu$, in the 1970's instead discovered 
unexpectedly large hadron production at high $p_T$, notably $\pi^0$, which provided a huge 
background of $e^{\pm}$ from internal and external conversions. Methods developed at the CERN 
ISR which led to the discovery of direct-single-$e^{\pm}$ in 1974, later determined to be from the 
semi-leptonic decay of charm which had not yet been discovered, were used by PHENIX at RHIC to make 
precision measurements of heavy quark production in p-p and Au+Au collisions, leading to the puzzle 
of apparent equal suppression of light and heavy quarks in the QGP. If the Higgs mechanism gives 
mass to gauge bosons but not to fermions, then a proposal that all 6 quarks are nearly massless in a 
QGP, which would resolve the puzzle, can not be excluded. This proposal can be tested with future 
measurements of heavy quark correlations in A+A collisions 

\end{abstract}
\section{Introduction}\label{sec:introduction}
   In September 1991 the PHENIX experiment was born from the rejection of the Dimuon, Oasis and TALES/Sparhc Letters of Intent (LoI) which were combined into a new collaboration to build ``a detector designed to study electrons and photons emerging from the QGP.'' Any collaborators ``primarily interested in hadron physics will be welcomed by the STAR collaboration which has been empowered to build a large TPC detector''~\cite{MelSLetter}. 
\section{PHENIX is not your father's solenoid collider detector}   
\begin{figure}[h]
\begin{center}
\includegraphics[width=0.95\linewidth]{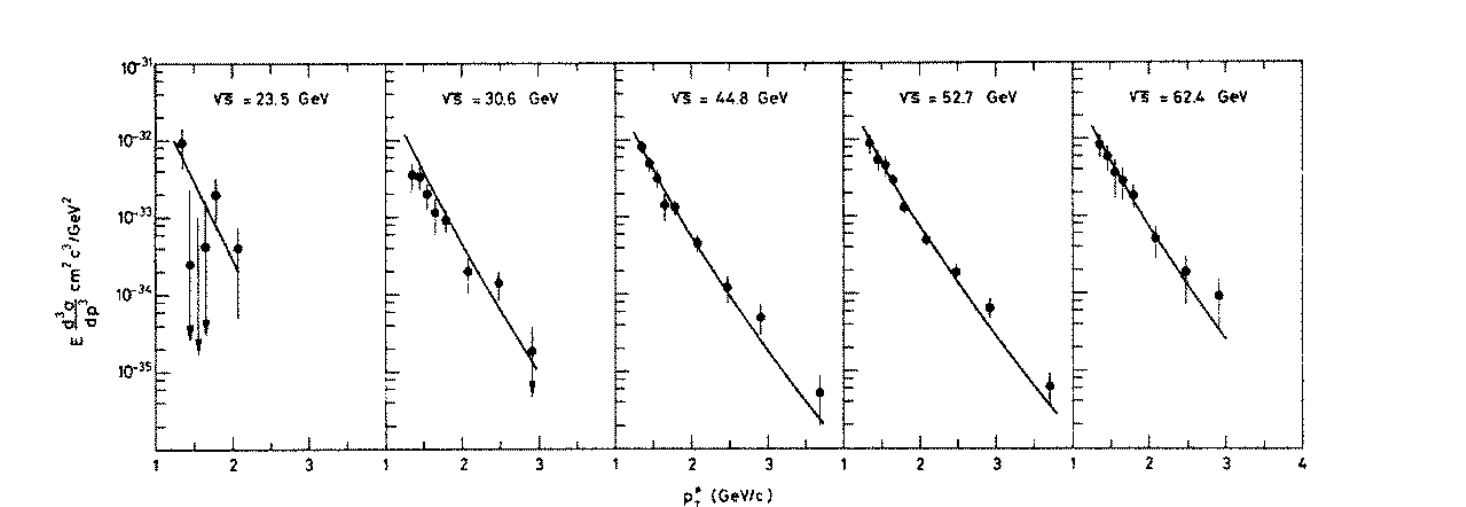}
\end{center}
\caption[]{Invariant cross sections at mid-rapidity for 5 values of $\sqrt{s}$ in p-p collisions at the CERN-ISR: $(e^+ + e^-)/2$ (points); $10^{-4}\times (\pi^+ +\pi^-)/2$ (lines)~\cite{CCRS}}
\label{fig:eroots}
\end{figure}

   The central spectrometer of PHENIX was designed to trigger on $J/\Psi\rightarrow e^+ e^-$ at rest, i.e. $p_T^{J/\Psi}\approx 0$ at mid-rapidity, based on experience from the CERN ISR  as proposed in the TALES~\cite{Tales} and later TALES/SPARHC~\cite{TS} LoI. The CCRS experiment at the CERN ISR had discovered direct-single $e^{\pm}$ in 1974~\cite{CCRS74} at a level of $10^{-4}$ of charged pion production at all 5 c.m. energies measured (Fig.~\ref{fig:eroots}).     
The same methods---i) $\geq 10^5$ 
charged hadron rejection; ii) minimum of material in the aperture to avoid external conversions; iii) 
zero magnetic Þeld on the axis to avoid de-correlating conversion pairs; iv) precision measurement of 
$\pi^0$ and $\eta$, the predominant background source; v) precision background determination in the 
direct-single-$e^{\pm}$ signal channel by adding external converter---were used in design of the highly non-conventional PHENIX mid-rapidity spectrometer, as I discussed in a previous winter workshop~\cite{MJTWWND04}. \begin{figure}[h]
\begin{minipage}{0.45\linewidth}
\includegraphics[width=\linewidth]{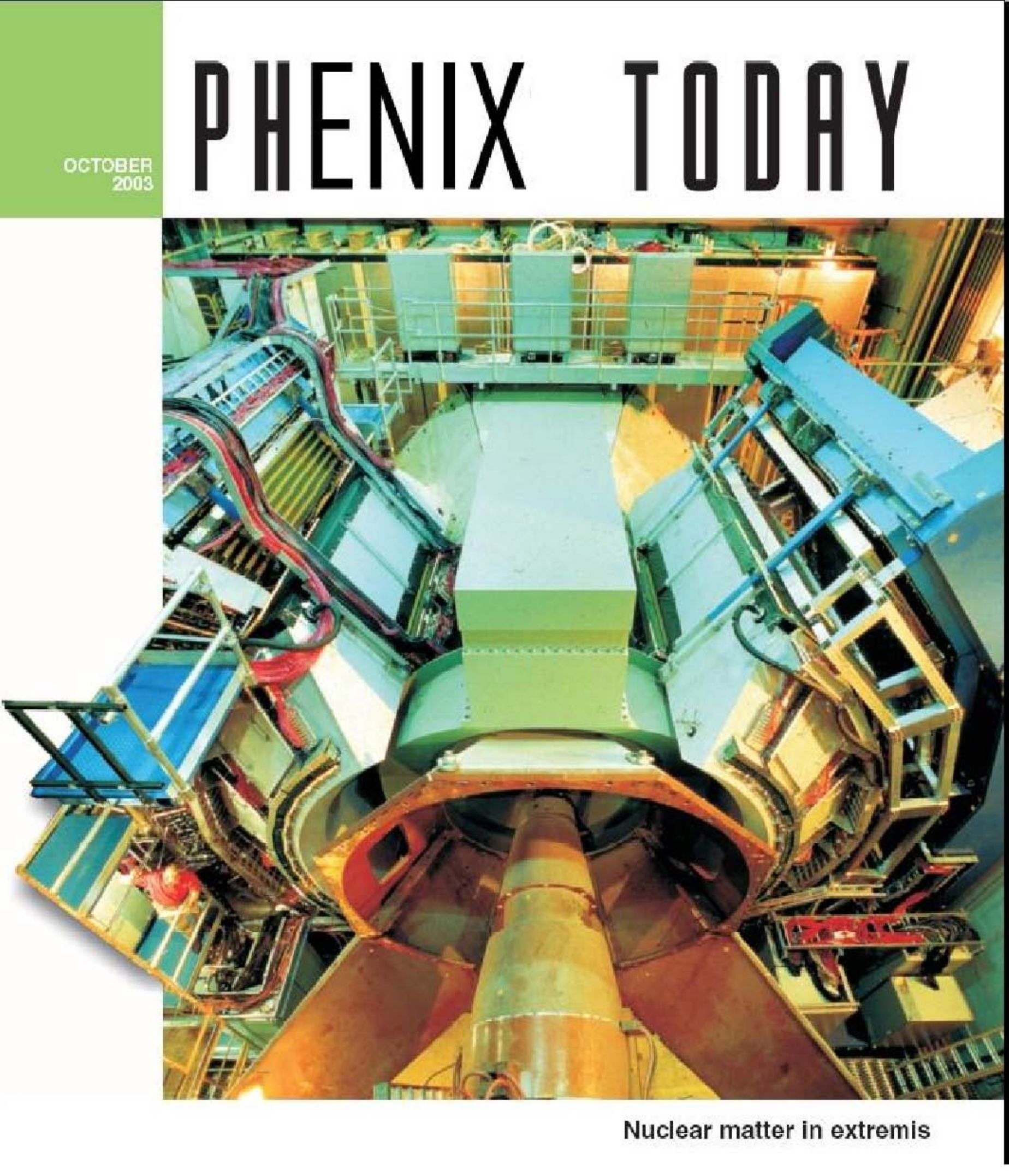}
\caption{\label{fig:PXtoday}PHENIX on cover of Physics Today, October 2003.}
\end{minipage}\hspace{2pc}%
\begin{minipage}{0.475\linewidth}The key detector elements in achieving these goals were a fine grained Electromagnetic Calorimeter and a Ring Imaging Cerenkov counter for use in achieving the charged hadron rejection and providing an electron trigger at level 1; and an axial field 
Helmholtz coil spectrometer magnet, rather than the typical solenoid detector generally used at colliders, with two sets of coils which could be run with opposite currents to achieve zero field on the axis. It is no surprise that PHENIX made the cover of Physics Today instead of the more conventional STAR detector because PHENIX is a special purpose detector. In fact when Jack Steinberger came to visit BNL and I was showing and explaining PHENIX to him, he frowned more than usual and said, ``Mike, is there a `real collider detector' at RHIC?" So I took him to STAR for a minute and he was happy.
\end{minipage}
\end{figure}
\section{A charming surprise}
\begin{figure}[!ht]
\begin{center} 
\begin{tabular}{cc}
\hspace*{-0.04\linewidth}\includegraphics*[width=0.53\linewidth]{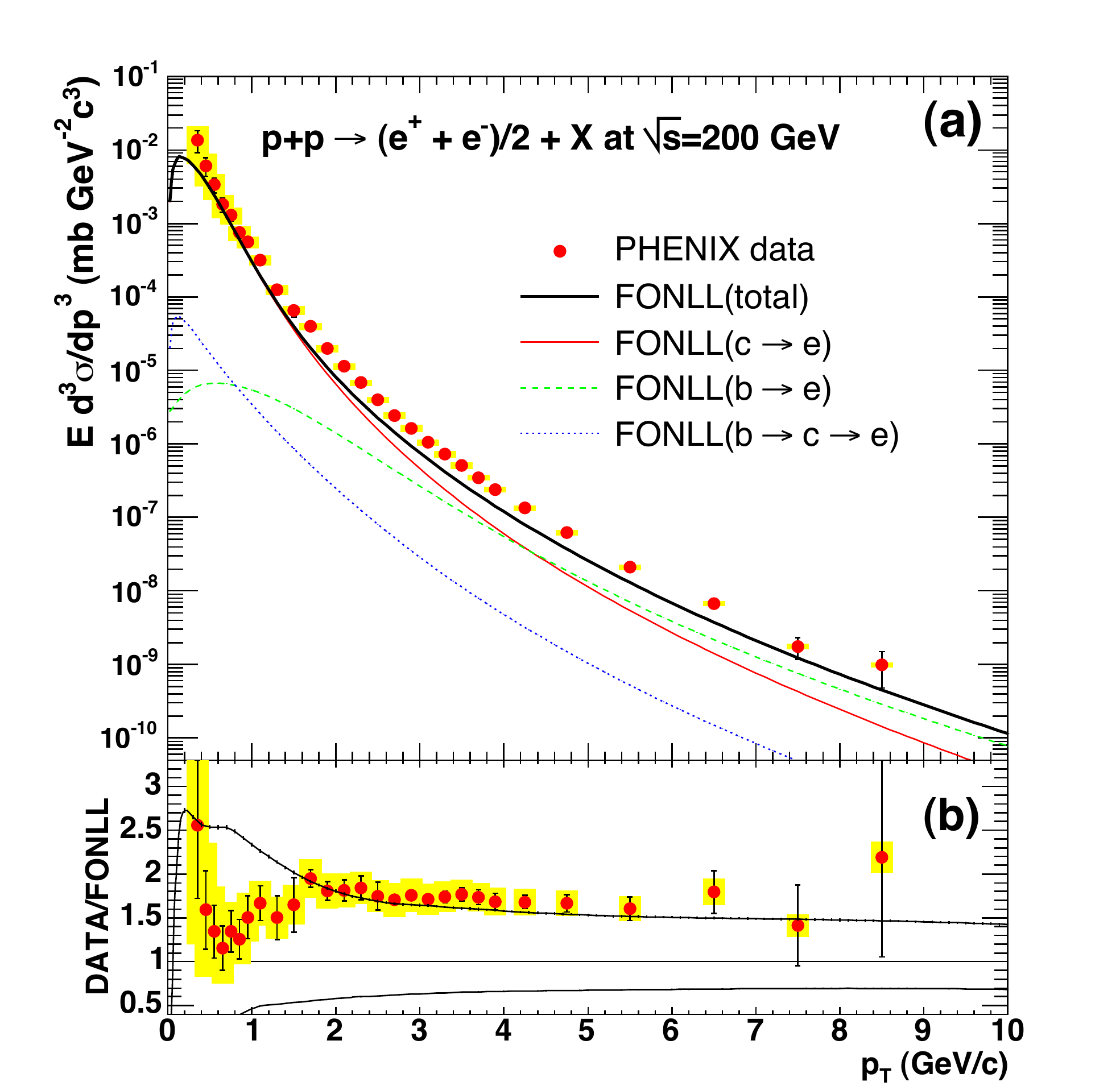} &  
\hspace*{-0.06\linewidth}\includegraphics[width=0.51\linewidth]{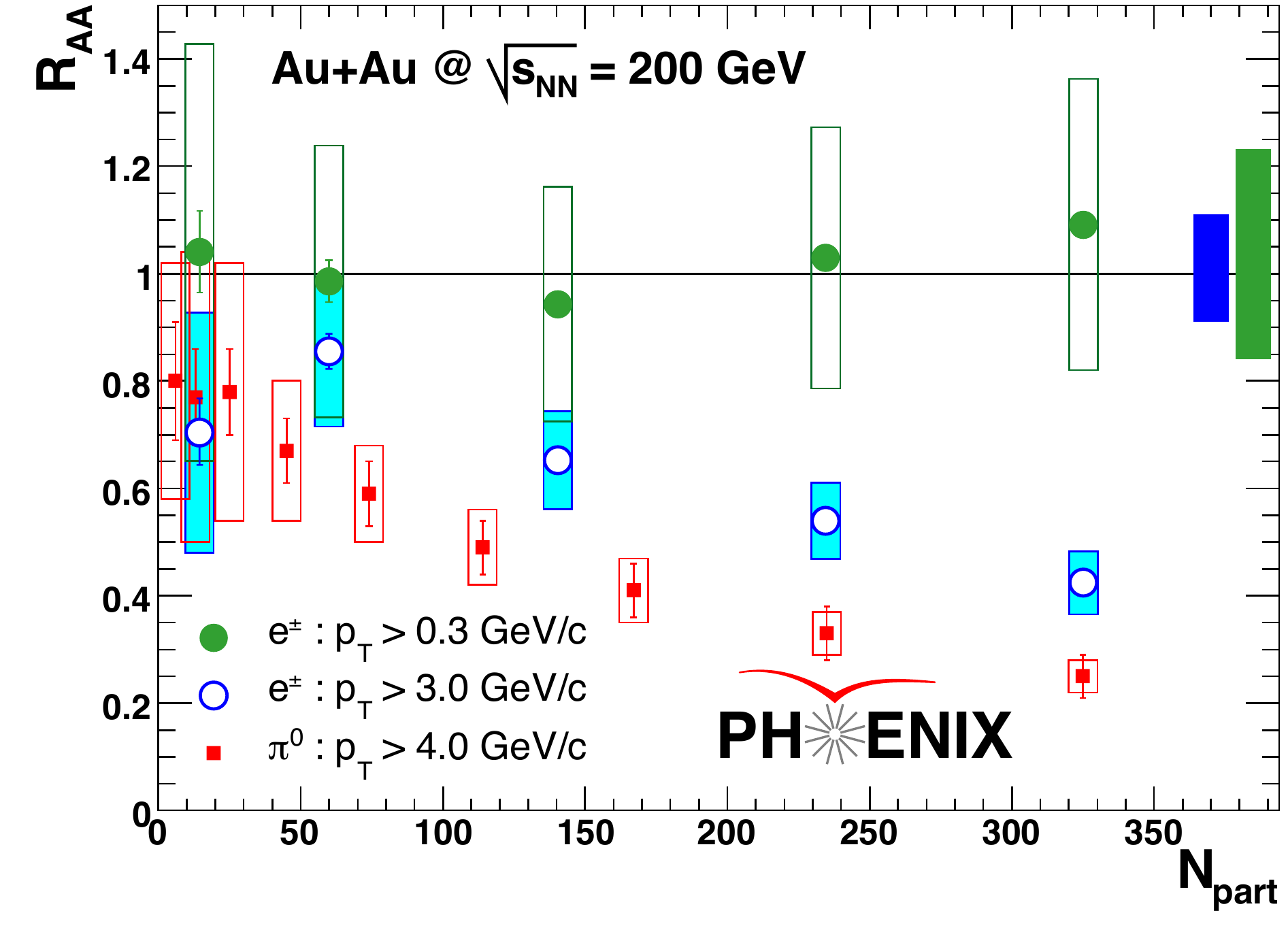} 
\end{tabular}
\end{center}\vspace*{-1.5pc}
\caption[]{a),b) (left) Invariant cross section of direct $e^{\pm}$ in p-p collisions ~\cite{PXcharmPRL97} compared to theoretical predictions from $c$ and $b$ quark semileptonic decay. c) (right) $R_{AA}$ as a function of centrality ($N_{\rm part}$) for the total yield of $e^{\pm}$ from charm ($p_T > 0.3$) GeV/c, compared to the suppression of the $e^{\pm}$ yield at large $p_T>3.0$ GeV/c which is comparable to that of $\pi^0$ with ($p_T>4$ GeV/c)~\cite{PXcharmAA06}}
\label{fig:f7}
\end{figure}

   We designed PHENIX specifically to be able to detect charm particles via direct-single $e^{\pm}$ since this went along naturally with $J/\Psi\rightarrow e^+ +e^-$ detection and since the single particle reaction avoided the huge combinatoric background in Au+Au collisions. We thought that the main purpose of open charm production, which corresponds to a hard-scale ($m_{c\bar{c}}\gsim 3$ GeV/c$^2$), would be a check of our centrality definition and $\mean{T_{AA}}$ calculation since the total production of $c$ quarks should follow point-like scaling. In fact, our first measurement supported this beautifully~\cite{PXcharmPRL94}. However, our subsequent measurements proved to be much more interesting and even more beautiful. Figure~\ref{fig:f7}a shows our direct-single-$e^{\pm}$ measurement in p-p collisions at $\sqrt{s}=200$ GeV~\cite{PXcharmPRL97} in agreement with a QCD calculation of $c$ and $b$ quarks as the source of the direct-single-$e^{\pm}$ (also called non-photonic $e^{\pm}$ at RHIC).   
The total yield of direct-single-$e^{\pm}$ for $p_T > 0.3$ GeV/c was taken as the yield of $c$-quarks in p-p and Au+Au collisions~\cite{PXcharmAA06}. The result, $R_{AA}=1$ as a function of centrality (Fig.~\ref{fig:f7}c), showed that the total $c-(\bar{c})$ production followed point-like scaling, as expected. The big surprise came at large $p_T$ where we found that the yield of direct-single-$e^{\pm}$ for $p_T>3$ GeV/c was suppressed nearly the same as the $\pi^0$ from light quark and gluon production. This strongly disfavors the QCD energy-loss explanation of jet-quenching because, naively, heavy quarks should radiate much less than light quarks and gluons in the medium; but opens up a whole range of new possibilities including string theory~\cite{egsee066}. 
\section{A History Lesson}
   ``Those who cannot remember the past are condemned to repeat it'', wrote George Santayna in 1905. Incredibly, the history of the discovery of direct-single-$e^{\pm}$ at the ISR by CCRS in 1974, which led to a spate of conflicting measurements by experiments not designed for the purpose, was repeated in 2005 at RHIC: STAR and PHENIX measurements of direct-$e^{\pm}$ in p-p collisions disagreed~\cite{STARePRL98}. I will not discuss this further except to insist that PHENIX was specifically designed for this measurement. However, I do think that it is worthwhile to review the history at the CERN-ISR where direct-single-$e^{\pm}$ were discovered before the $J/\Psi$ or charm particles were known and presented at the ICHEP in London in July 1974~\cite{CCRSLondon74}.

\begin{figure}[h]
\begin{center}
\includegraphics[width=0.75\linewidth]{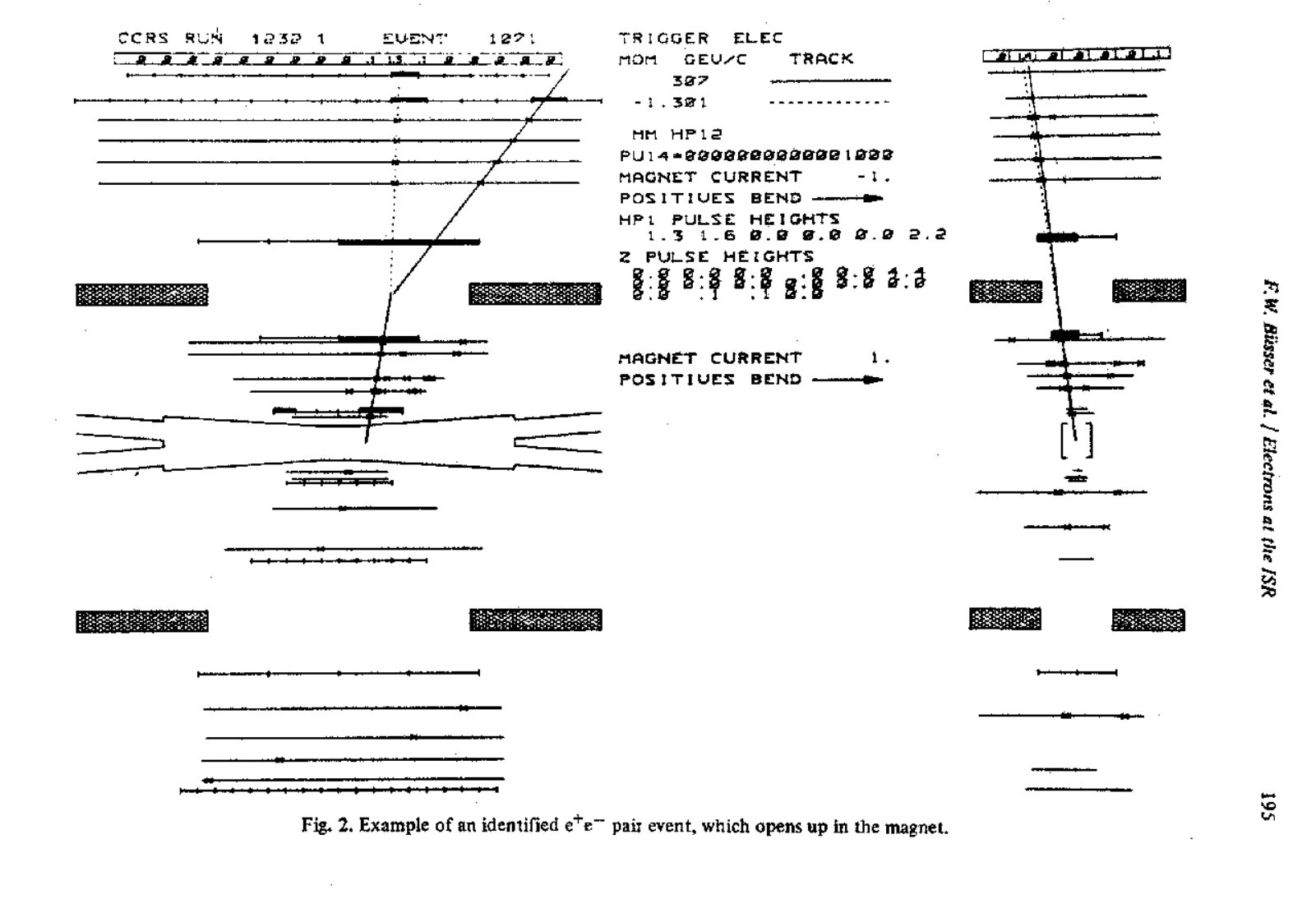}
\end{center}
\caption[]{CCRS~\cite{CCRS} identified $e^+ e^-$ pair which opens up in the magnet}
\label{fig:ccrsconv}
\end{figure}

    The first interpretation of the CCRS discovery was by Farrar and Frautschi~\cite{Glennys76} who proposed that the direct-single-$e^{\pm}$ were due to the internal conversion of direct photons with a ratio $\gamma/\pi^0\sim$10-20\%. CCRS was able to cleanly detect both external and internal conversions since there was zero magnetic field on the axis (Fig.~\ref{fig:ccrsconv}) and set limits excluding this explanation~\cite{CCRS}. 

\begin{figure}[h]
   \begin{center}
\includegraphics[width=0.42\linewidth]{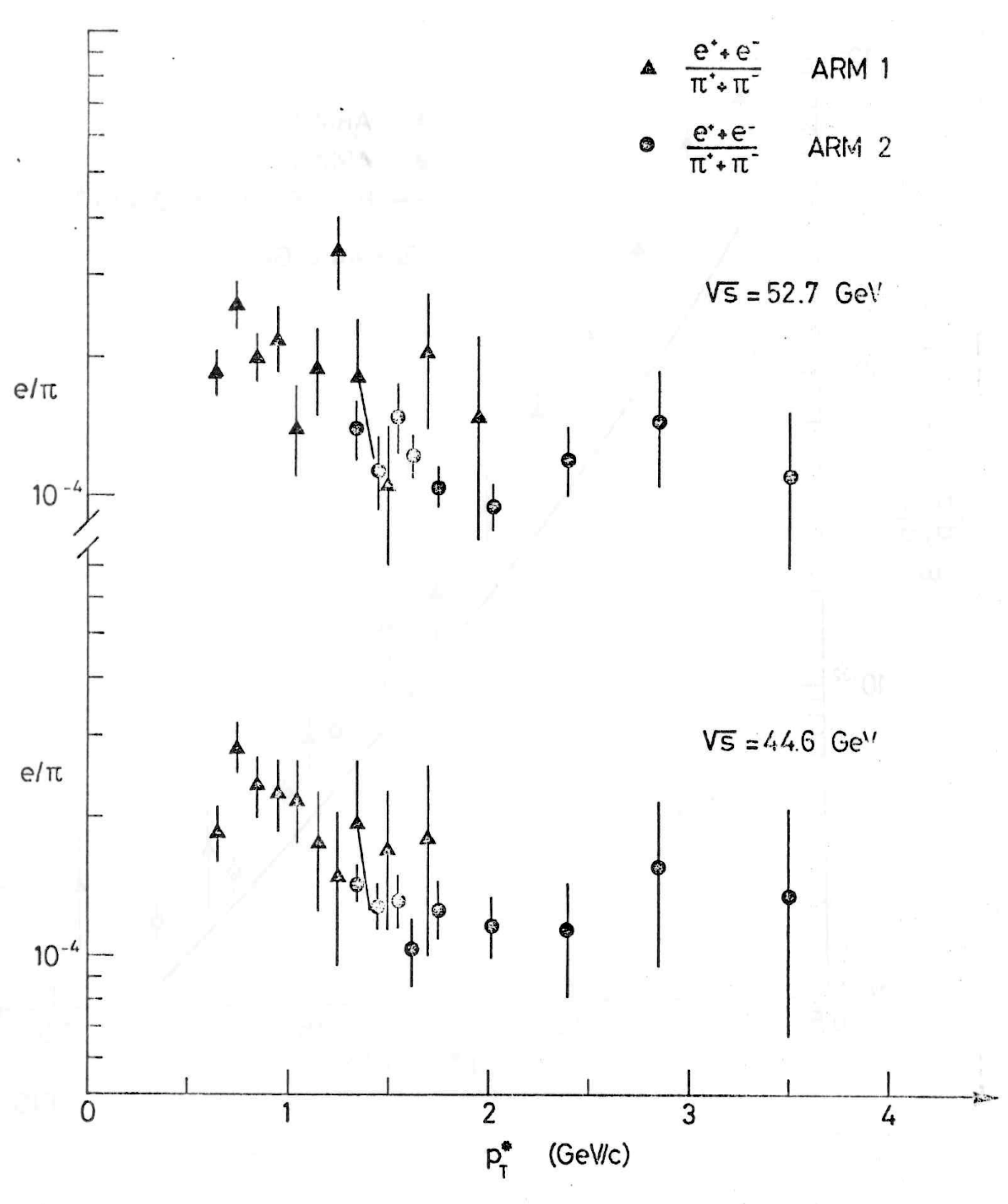}
\includegraphics[width=0.50\linewidth]{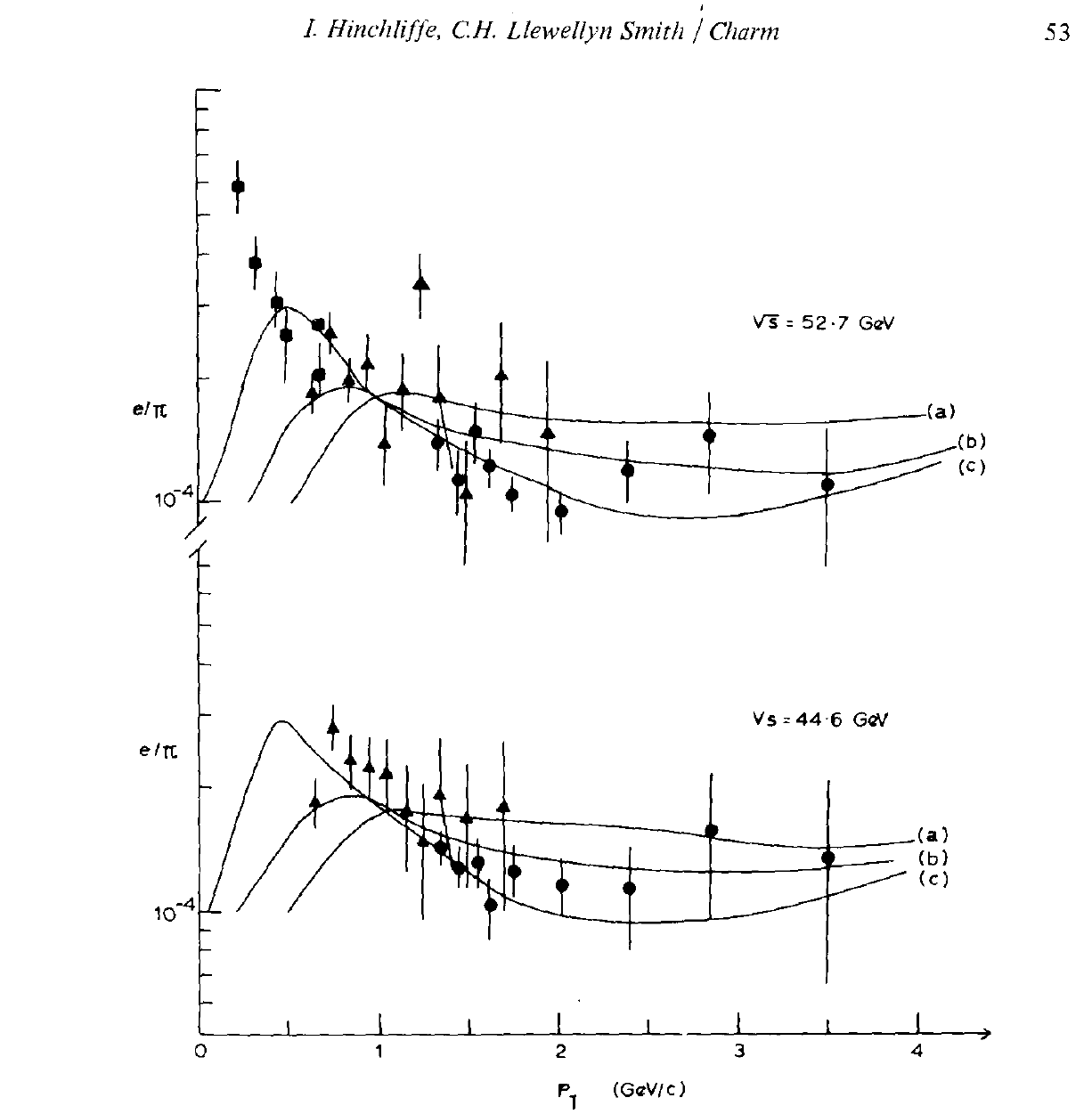}
\end{center}\vspace*{-0.25in}
\caption[]{a)(left) CCRS $e/\pi$ in two spectrometer arms at two values of $\sqrt{s}$~\cite{CCRS75} b)(right) same data with prediction, see Ref.~\cite{HLLS} for details
\label{fig:HLLS}}\vspace*{-1.0pc}
\end{figure}

\begin{figure}[h]
   \begin{center}
\includegraphics[width=0.56\linewidth]{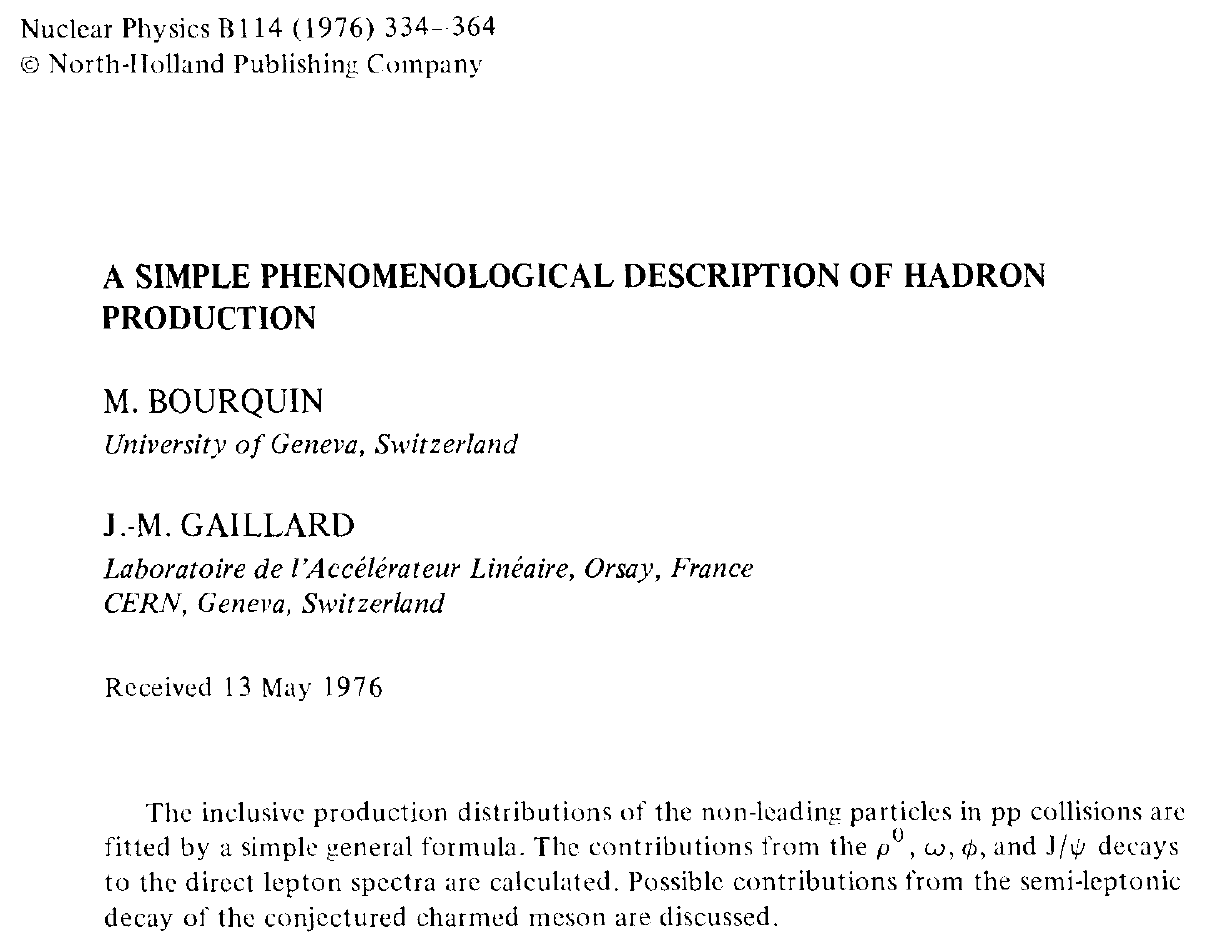}
\includegraphics[width=0.35\linewidth]{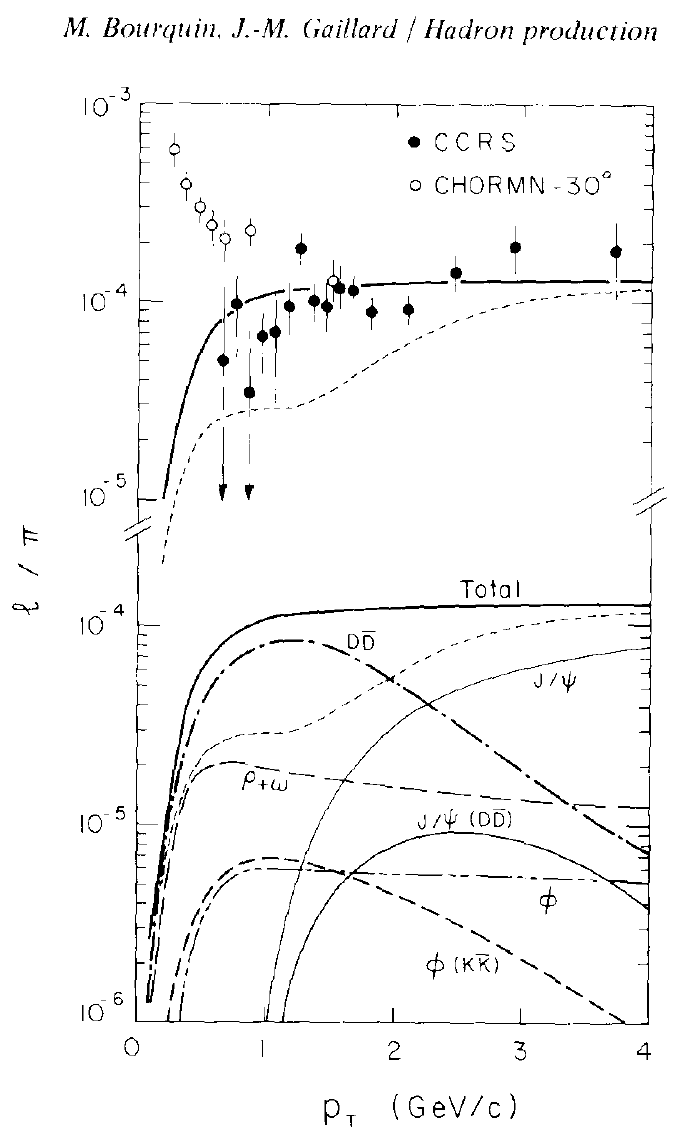}
\end{center}\vspace*{-0.25in}
\caption[]{(left) Bourquin-Gaillard title page~\cite{BG76}; (right) Contribution of a cocktail of decays to the $e/\pi$ ratio.   
\label{fig:BourqGaill}}
\end{figure}

\begin{figure}[t]
\begin{minipage}{0.45\linewidth}
\includegraphics[width=\linewidth]{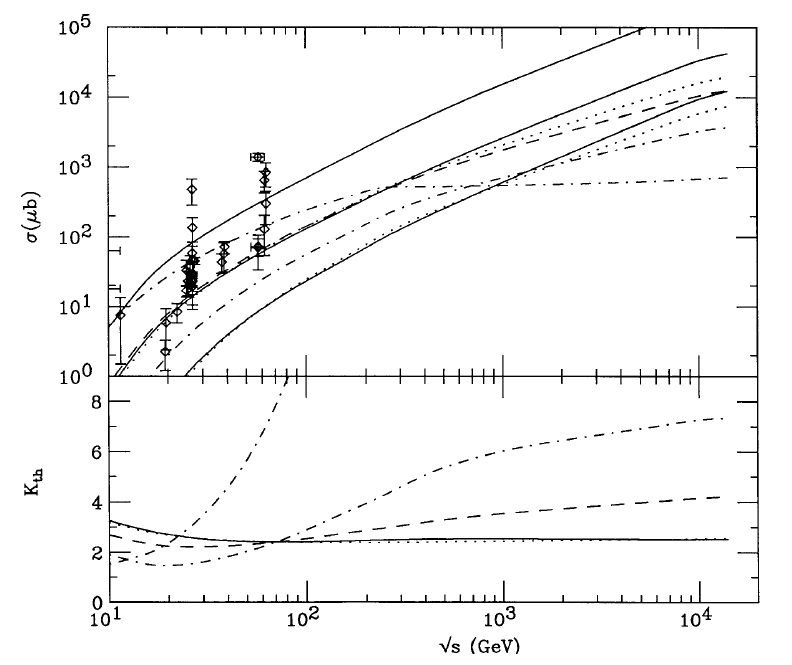}
\caption{\label{fig:RamonaV}Predictions and measurement of open charm cross section~\cite{Vogt96}}
\end{minipage}\hspace{1pc}%
\begin{minipage}{0.5\linewidth}
\hspace*{-0.015\linewidth}\includegraphics[width=1.05\linewidth]{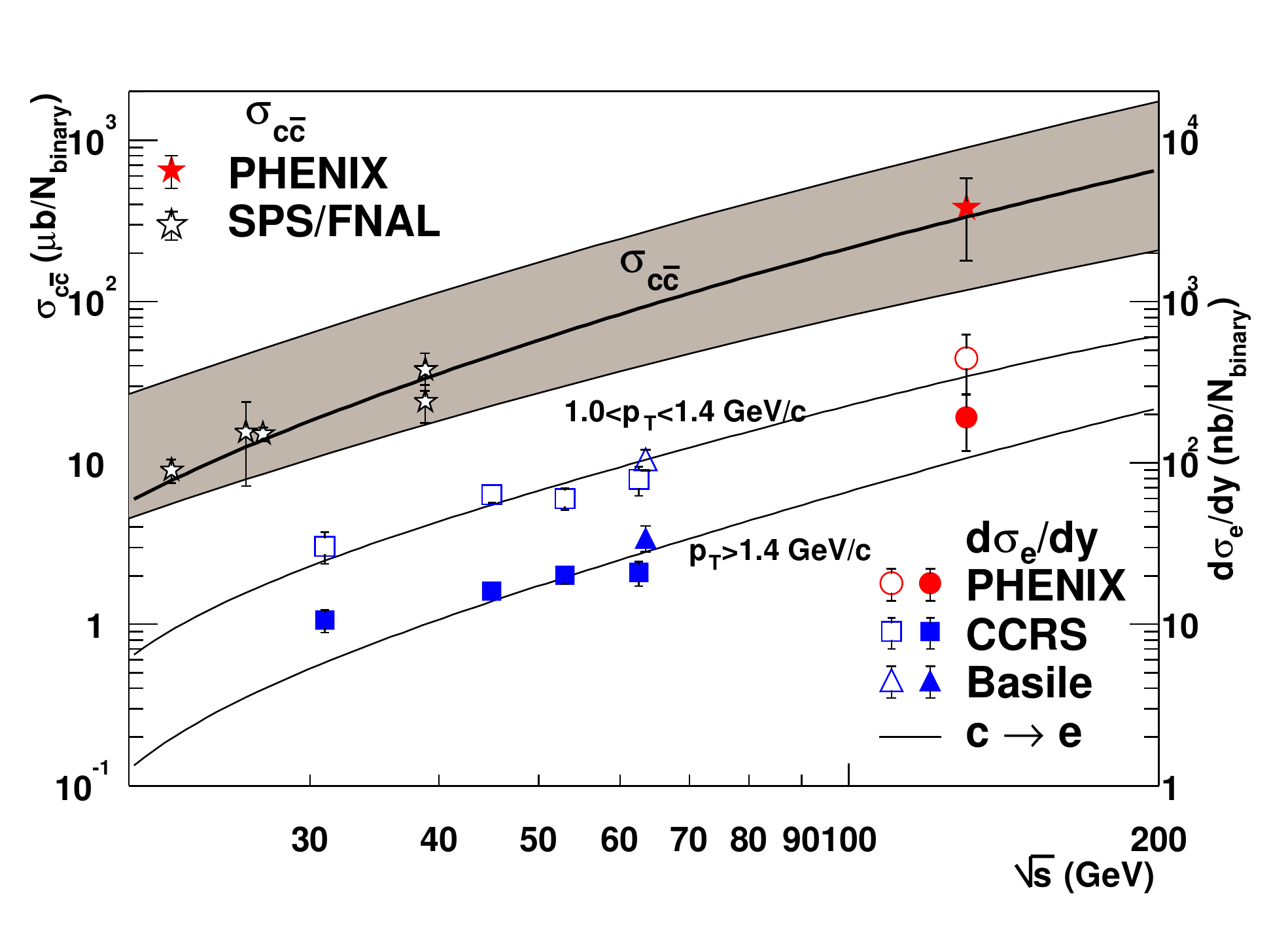}
\caption{\label{fig:prl88f4}Predictions and measurements~\cite{PXPRL88} of charm cross section $\sigma_{c\bar{c}}$}
\end{minipage} 
\end{figure}

	The first correct explanation of the CCRS direct-single-$e^{\pm}$ (prompt leptons) was given by Hinchliffe and Llewellyn-Smith~\cite{HLLS} as due to semi-leptonic decay of charm particles. Open charm was discussed at the 1975 Lepton-Photon conference at SLAC, the first major conference after the discovery of the $J/\Psi$ and $\Psi'$, and the paper submitted to Physical Review Letters in June 1975, but was not published until August 1976~\cite{OpenCharm}. The CCRS data submitted to the SLAC conference~\cite{CCRS75} and the prediction from charm decay~\cite{HLLS} are shown in Fig.~\ref{fig:HLLS}.
A similar explanation was offered by two experimentalists, Maruice Bourquin and Jean-Marc Gaillard~\cite{BG76}, who compared the measured $e/\pi$ ratios to a cocktail of all known leptonic decays including {\em ``Possible contributions from the conjectured charm meson...''} (Fig.~\ref{fig:BourqGaill}). The usage of ``conjectured'' charm is notable due to the delayed publication of reference~\cite{OpenCharm}.
Another notable point about both papers~\cite{HLLS,BG76} is that neither could fit the data points 
\begin{figure}[h]
\begin{center}
\includegraphics[width=0.85\linewidth]{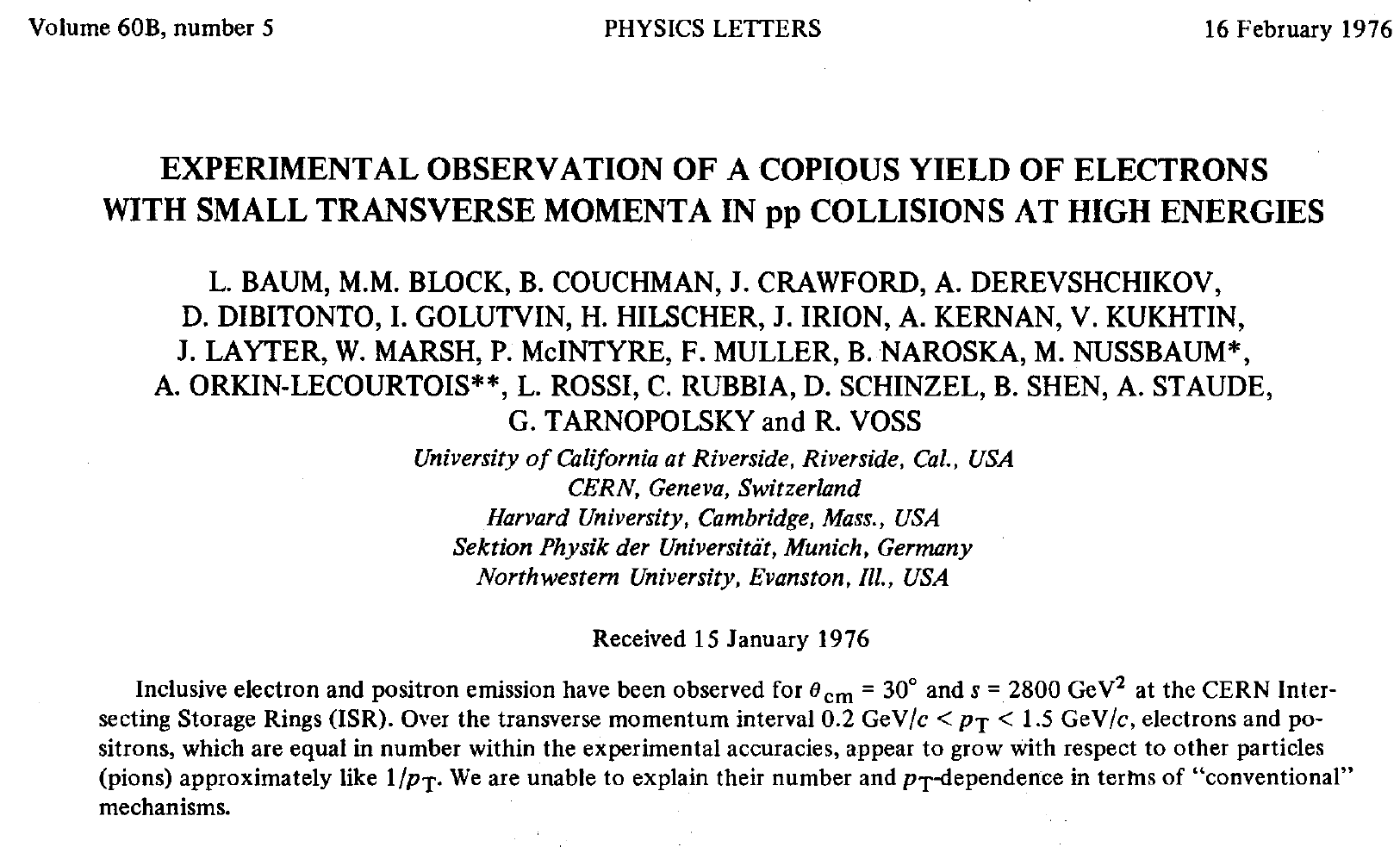}
\end{center}
\caption[]{One of the stars of 1976, measurement at 30$^\circ$ and 0.2 GeV/c $<p_T <$  1.5 GeV/c}
\label{fig:chormn}
\end{figure}
for $p_T<1$ GeV/c from the CHORMN measurement~\cite{chormn} (Fig.~\ref{fig:chormn}). The important point here is that many experiments not designed for the purpose wanted to get into the prompt-lepton act and this is only one example. See Ref.~\cite{BG76} for more details. This led to the unfortunate  situation as exemplified in Fig.~\ref{fig:RamonaV} from a typical paper~\cite{Vogt96} about charm c. 1990 which didn't even give citations for 
the many disagreeing ISR publications of the direct-single-$e^{\pm}$ from charm which were essentially ridiculed or, in the case of the original CCRS discovery~\cite{CCRS75}, ignored~\cite{Tavernier} because it was published before either charm or the $J/\Psi$ were discovered so there is no reference to the word ``charm'' in the publication. A fairer comparison of the ISR and fixed target measurements~\cite{Appel92} of open charm is given in Fig.~\ref{fig:prl88f4} from the first measurement of direct-single-$e^{\pm}$ at RHIC by PHENIX~\cite{PXPRL88}.
   \begin{figure}[h]
   \begin{center}
\includegraphics[width=0.46\linewidth]{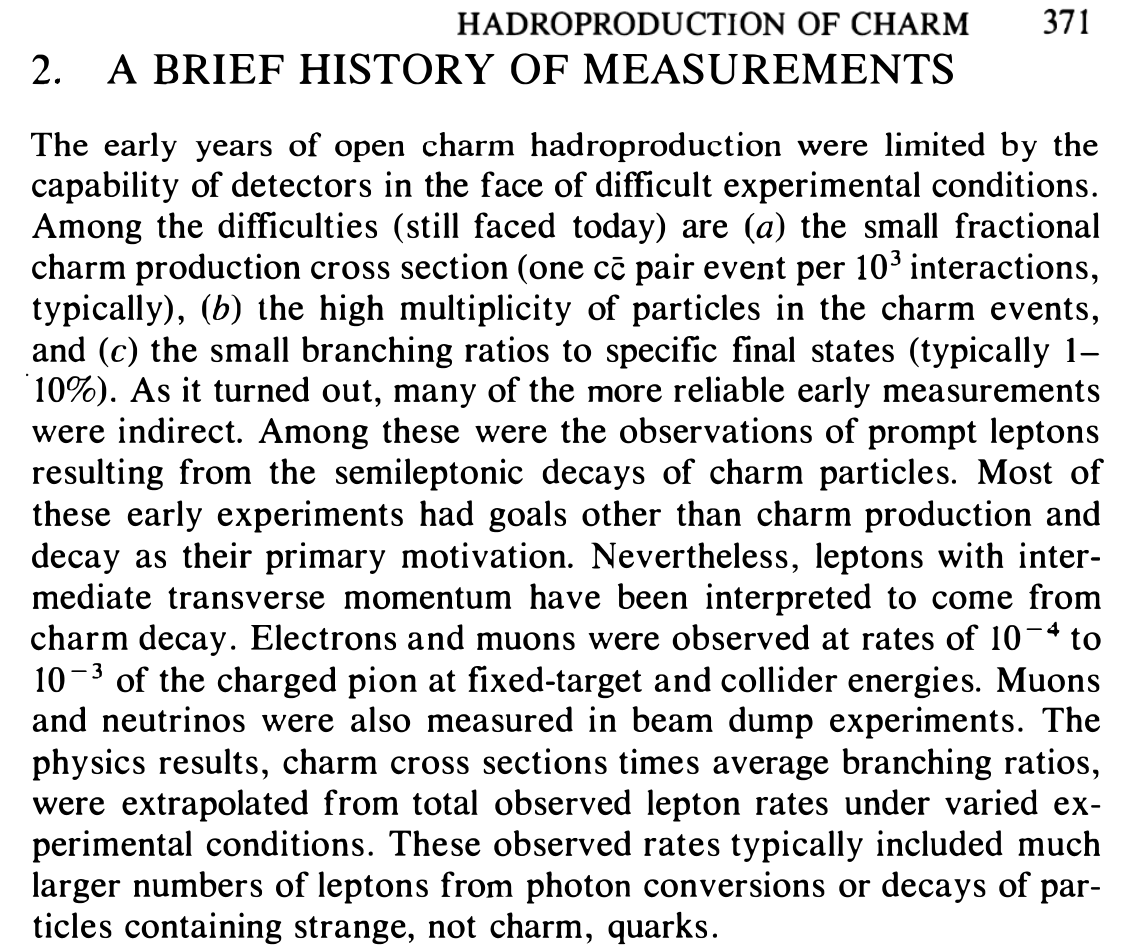}
\includegraphics[width=0.45\linewidth]{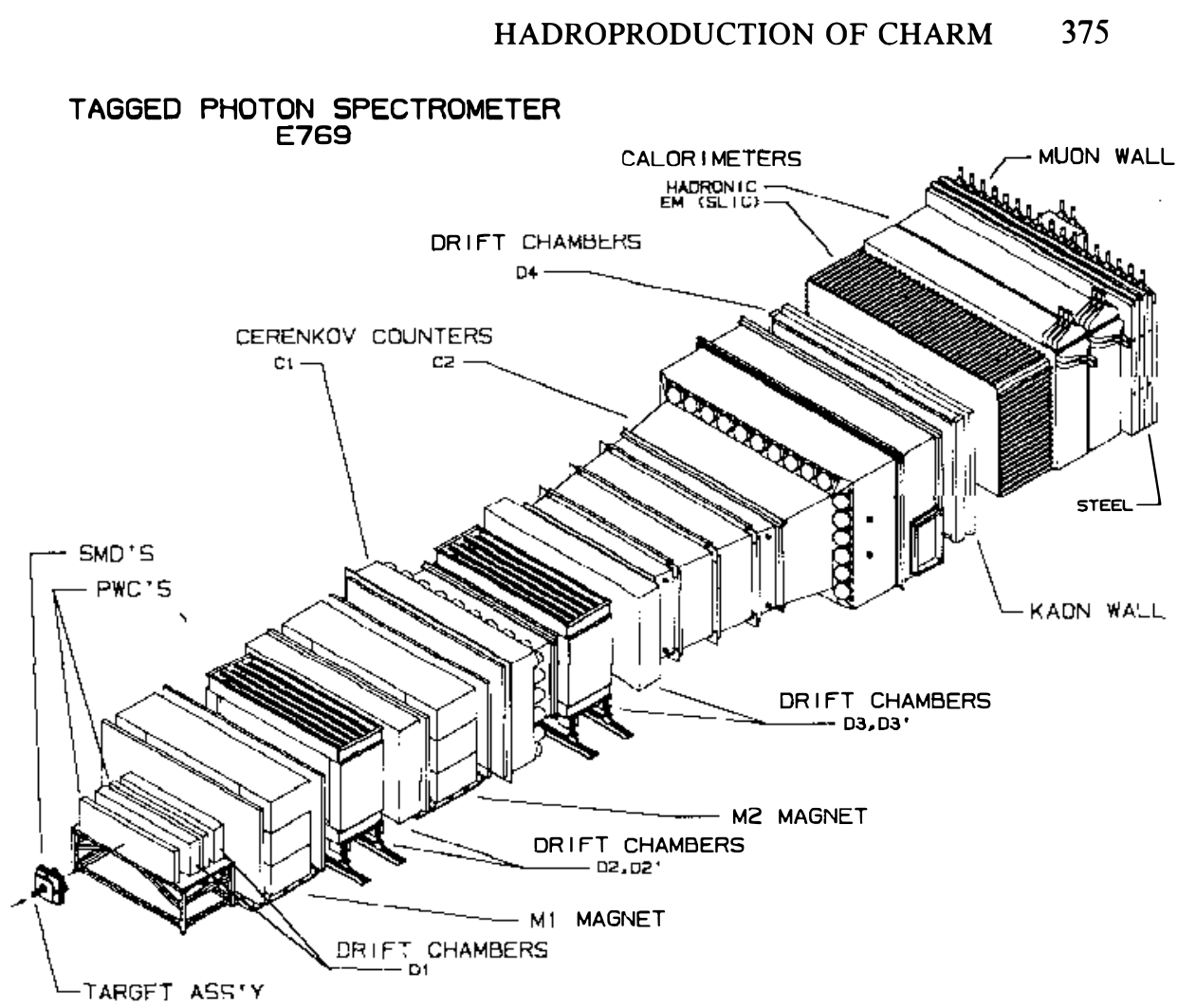}
\end{center}\vspace*{-0.25in}
\caption[]{a)(left) Text and b)(right) figure from Jeff Appel's review article~\cite{Appel92}.    
\label{fig:Appel}}
\end{figure}
The fixed target measurements, which like to claim the discovery of hadroproduction of charm, all used Silicon Vertex Detectors, which is the first tiny element (SMD), actually an upgrade, in an otherwise giant experiment (Fig.~\ref{fig:Appel}b). At least, Jeff Appel in his review article~\cite{Appel92} (Fig.~\ref{fig:Appel}a) did acknowledge the much earlier prompt lepton results.  

\subsection{Contemporary (c. 1977) controversies}
    To be fair, the direct-single-$e^{\pm}$ measurements as observations of semi-leptonic decays of open charm were not universally accepted by the experimental community in the late 1970's; and there were heavy hitters on both sides of the argument. First of all, after the discovery of the $J/\Psi$ in November 1974 and, in particular, with the near miss of this discovery at the CERN-ISR, there was concern that the $J/\Psi\rightarrow e^+ +e^-$ decay could be the source of the single $e^{\pm}$.  
\begin{figure}[t]
\begin{center}
\begin{tabular}{ccc}
\hspace*{-0.02\linewidth}\includegraphics[width=0.31\linewidth]{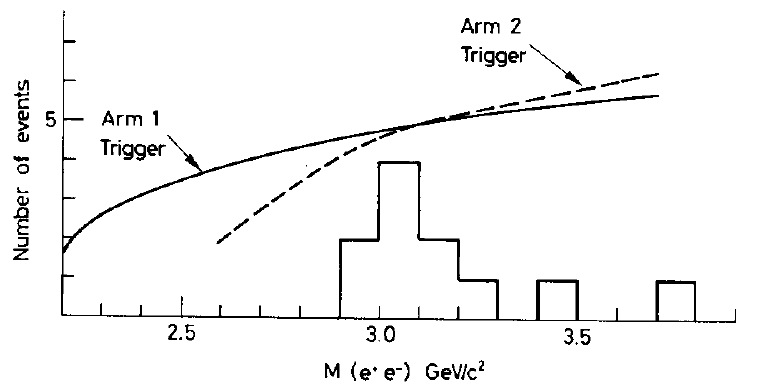}&
\hspace*{-0.04\linewidth}\includegraphics[width=0.33\linewidth]{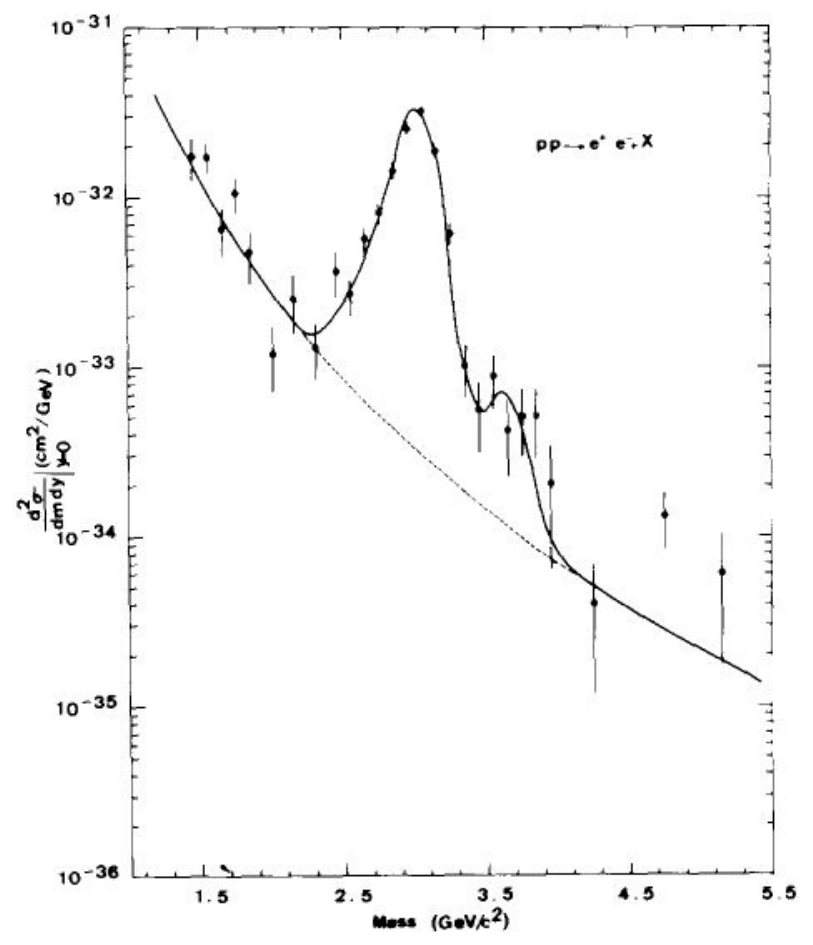}&
\hspace*{-0.02\linewidth}\includegraphics[width=0.33\linewidth]{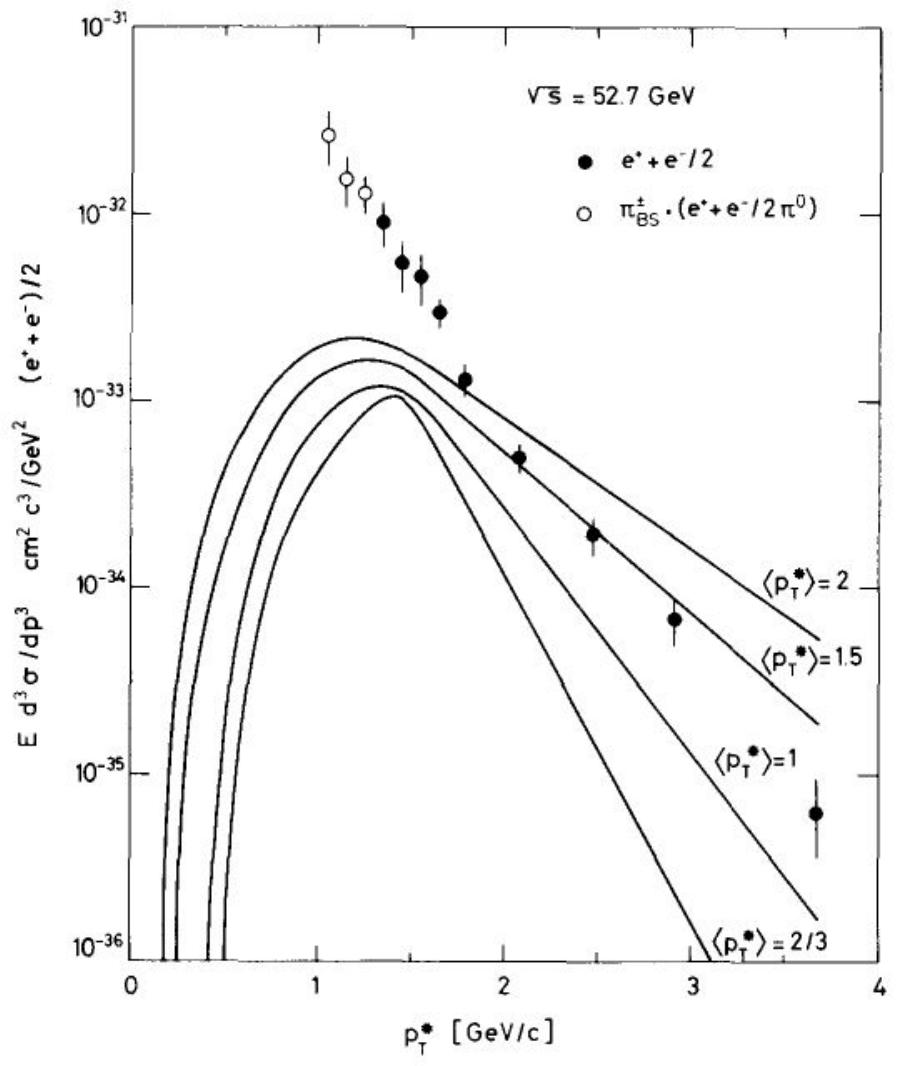} 
\end{tabular}
\end{center}
\caption[]{a)(left) First $J/\Psi$ at ISR~\cite{CCRSPLB56}; b) (center) Best $d\sigma_{ee}/dm_{ee} dy|_{y=0}$~\cite{Clark}; c)(right) direct-$e^{\pm}$ data at $\sqrt{s}=52.7$ GeV (Fig.~\ref{fig:eroots}) with calculated $e^{\pm}$ spectrum for $J/\Psi$ for several values of $\mean{p_T}$~\cite{CCRS}. }
\label{fig:I3}
\end{figure}
Fortunately, CCRS quickly demonstrated that the $J/\Psi$ was not the source of the single-$e^{\pm}$ (Fig.~\ref{fig:I3}).  Fig.~\ref{fig:I3}a~\cite{CCRSPLB56} shows the first $J/\Psi$ at the ISR~\cite{CCRSPLB56}, Fig.~\ref{fig:I3}b shows the best $J/\Psi$ measurement at the ISR~\cite{Clark} while Fig.~\ref{fig:I3}c~\cite{CCRS} shows that the direct electrons (Fig.~\ref{fig:eroots}) are not the result of $J/\Psi$ decay since $\mean{ p_T}=1.1\pm 0.05$ GeV/c~\cite{Clark}.  

\begin{figure}[h]
   \begin{center}
\includegraphics[width=0.75\linewidth]{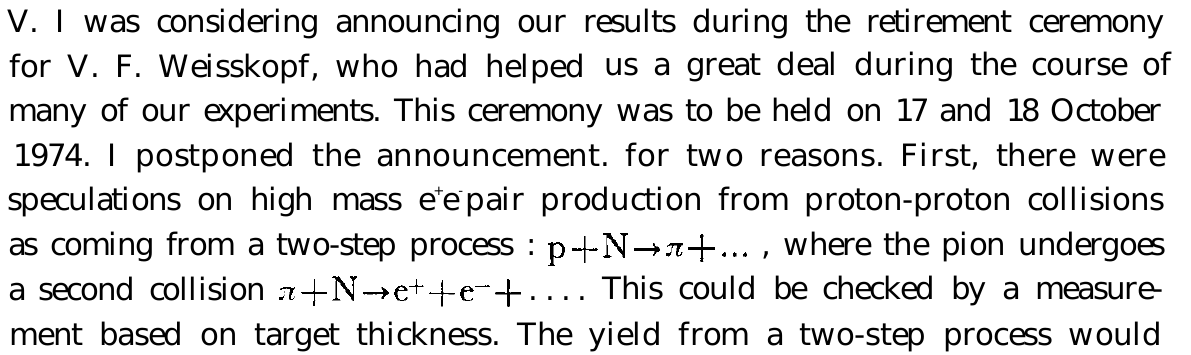}
\includegraphics[width=0.75\linewidth]{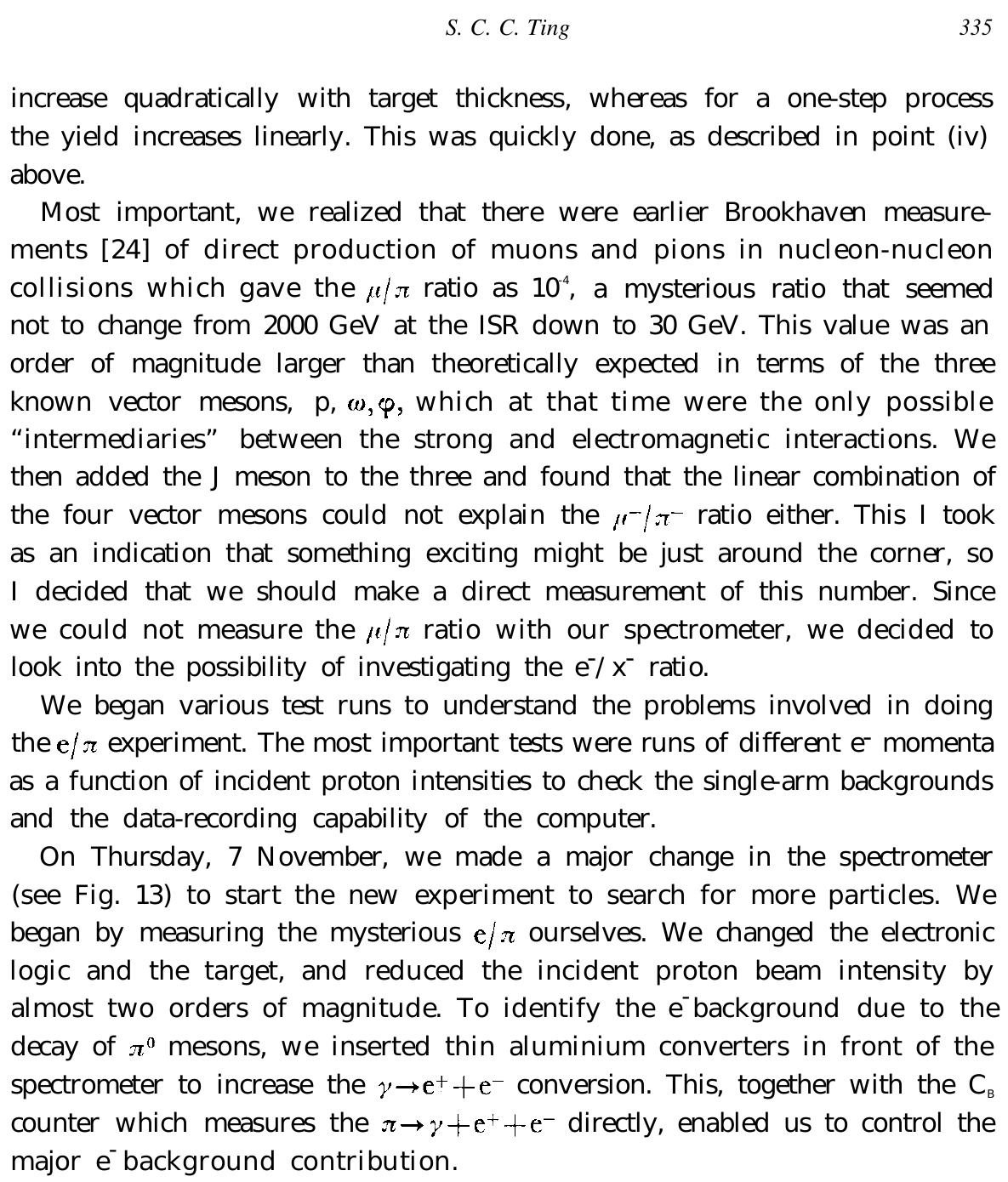}
\end{center}\vspace*{-0.25in}
\caption[]{Excerpt from Ting Nobel Lecture~\cite{TingNobel}
\label{fig:Ting}}
\end{figure}

Sam Ting, in his Nobel Lecture~\cite{TingNobel} (Fig.~\ref{fig:Ting}) also noted that the $J$ meson could not explain the prompt leptons and indicated that he actually delayed announcing the $J$ discovery at Viki Weisskopf's retirement ceremony in mid-October 1974 in order to investigate the prompt leptons in his AGS experiment. 

\begin{figure}[h]
   \begin{center}
\includegraphics[width=0.75\linewidth]{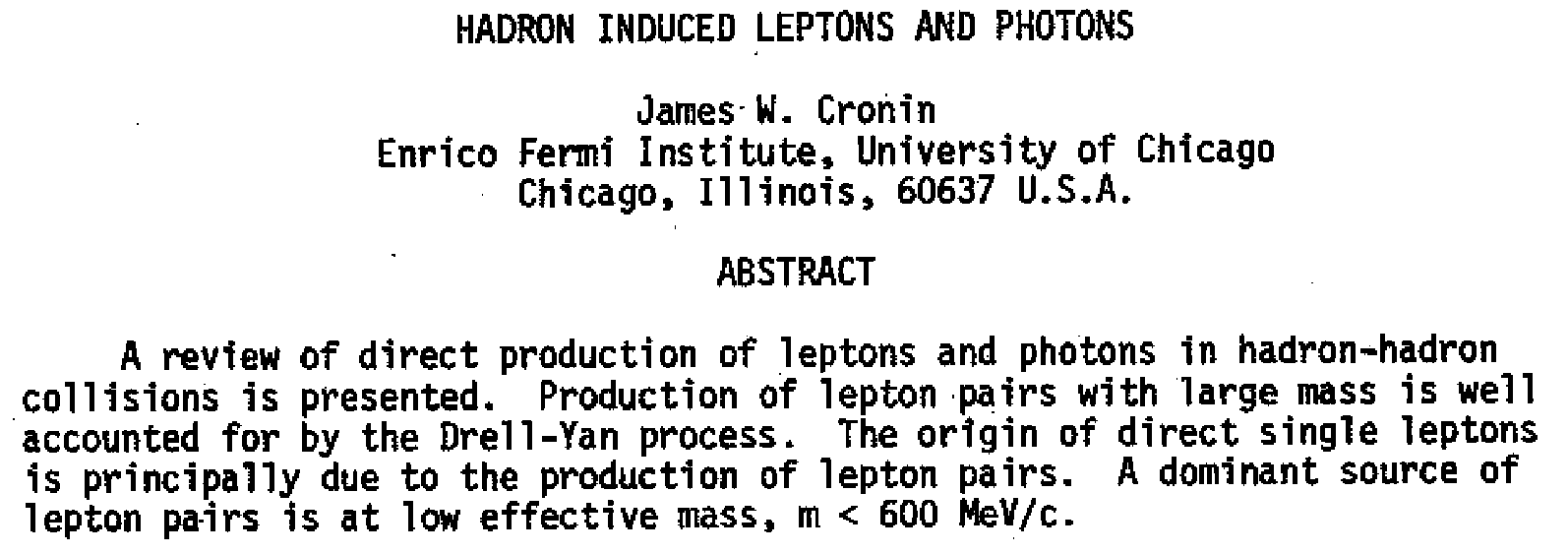}
\includegraphics[width=0.75\linewidth]{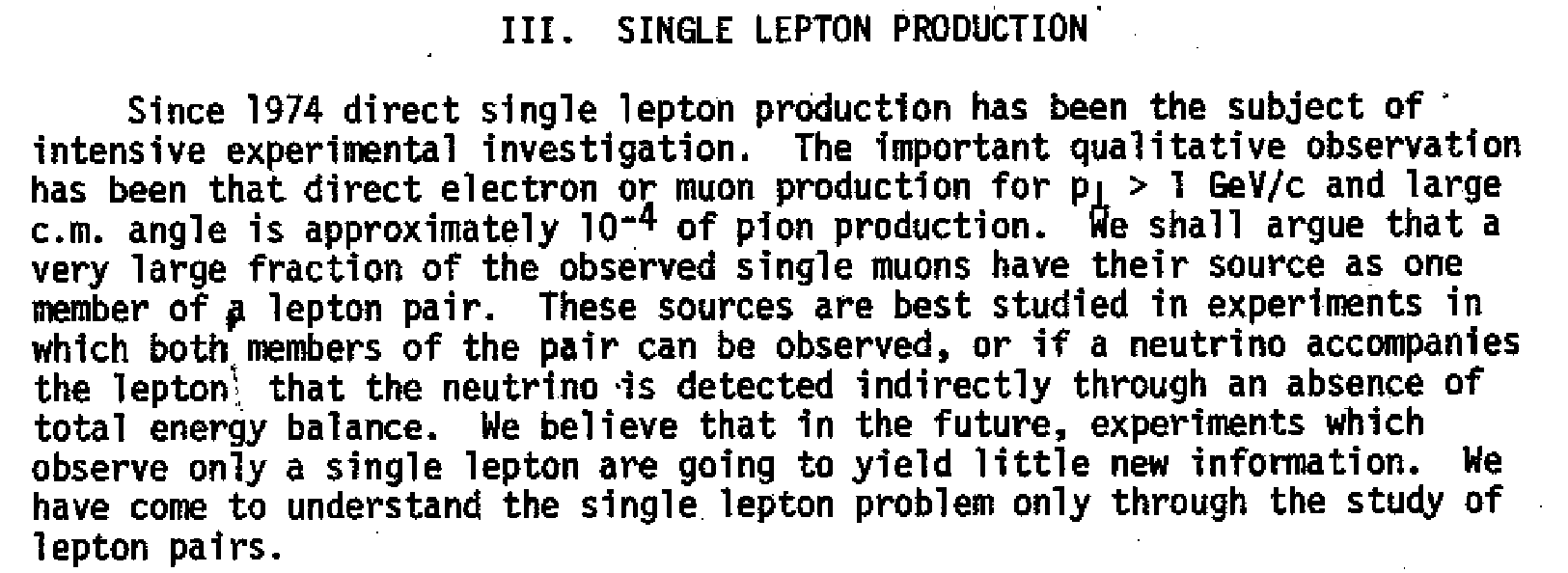}
\includegraphics[width=0.75\linewidth]{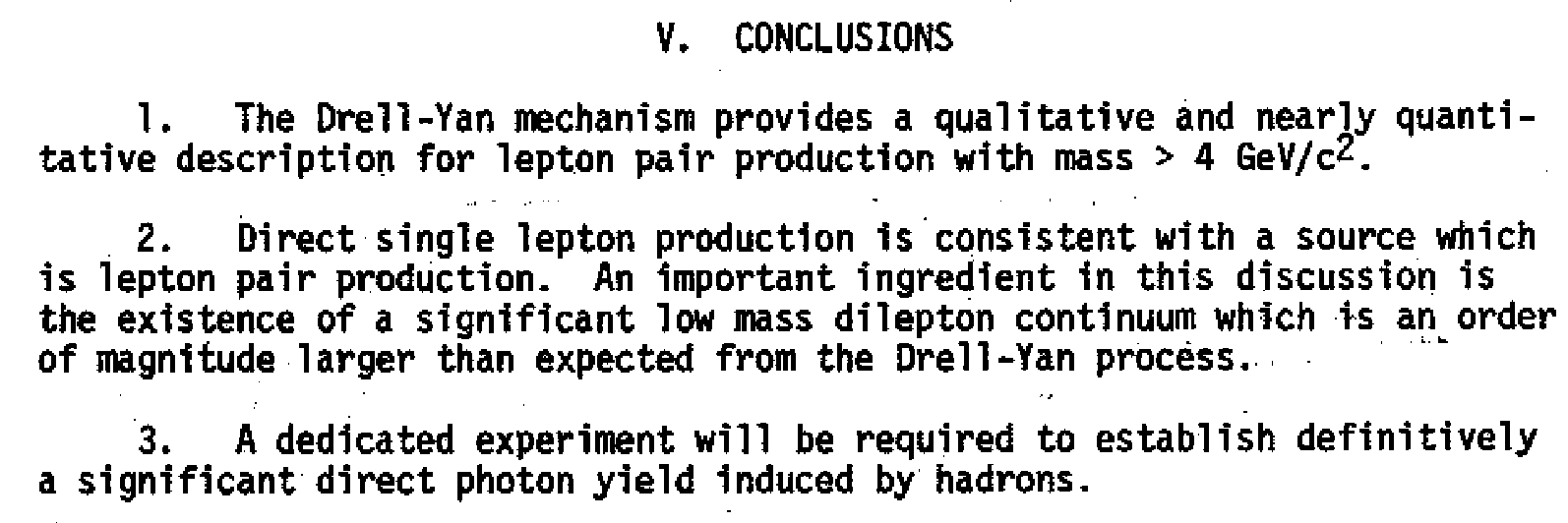}
\includegraphics[width=0.375\linewidth]{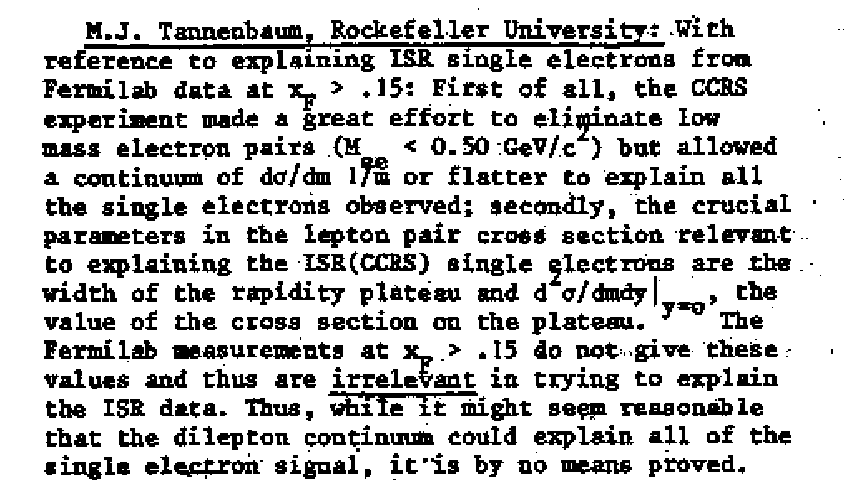}
\includegraphics[width=0.375\linewidth]{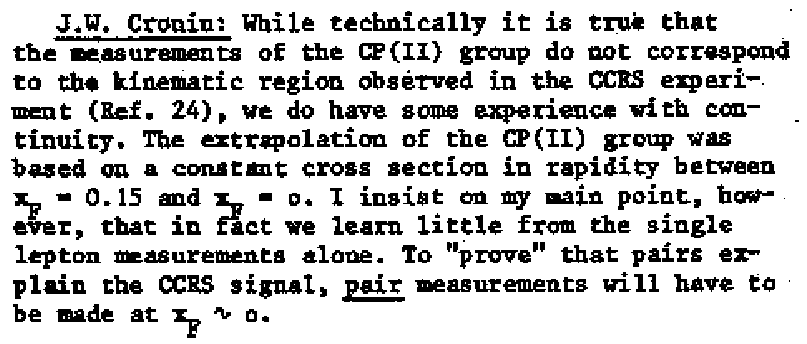}
\end{center}\vspace*{-0.25in}
\caption[]{Excerpt from Cronin talk and discussion at Lepton-Photon 1977~\cite{Cronin77} 
\label{fig:Cronin}}
\end{figure}

On the other side of this argument was Jim Cronin, another Nobel Laureate (but incorrect on this issue), who prominently claimed in his plenary talk at the 1977 Lepton Photon Symposium in Hamburg~\cite{Cronin77} that ``The origin of direct single leptons is principally due to the production of lepton pairs'' (Fig.~\ref{fig:Cronin}).
Also, as one can see from the discussion, a young (at the time) Associate Professor from the Rockefeller University vehemently disagreed with Cronin's explanation. In private, I also said, with evidence~\cite{JaneErratum}, that ``people who try to measure prompt leptons become the world's experts  on $\eta$ Dalitz decay''.  

\section{Back to the ``Heavy Quark Suppression Crisis'' at RHIC}
\begin{figure}[!ht]
\begin{center} 
\begin{tabular}{cc}
\hspace*{-0.02\linewidth}\includegraphics*[width=0.51\linewidth]{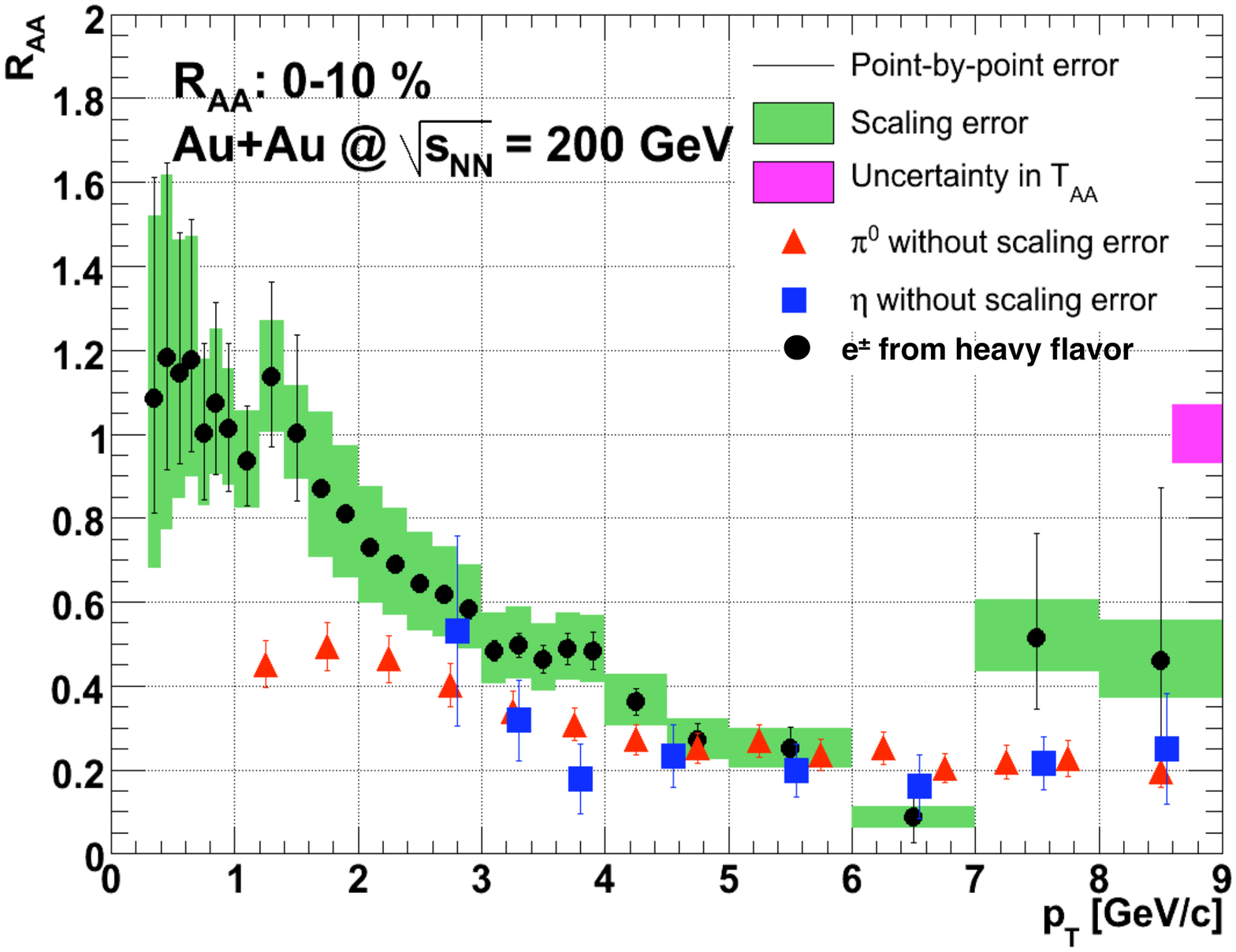} &  
\hspace*{-0.02\linewidth}\includegraphics[width=0.50\linewidth]{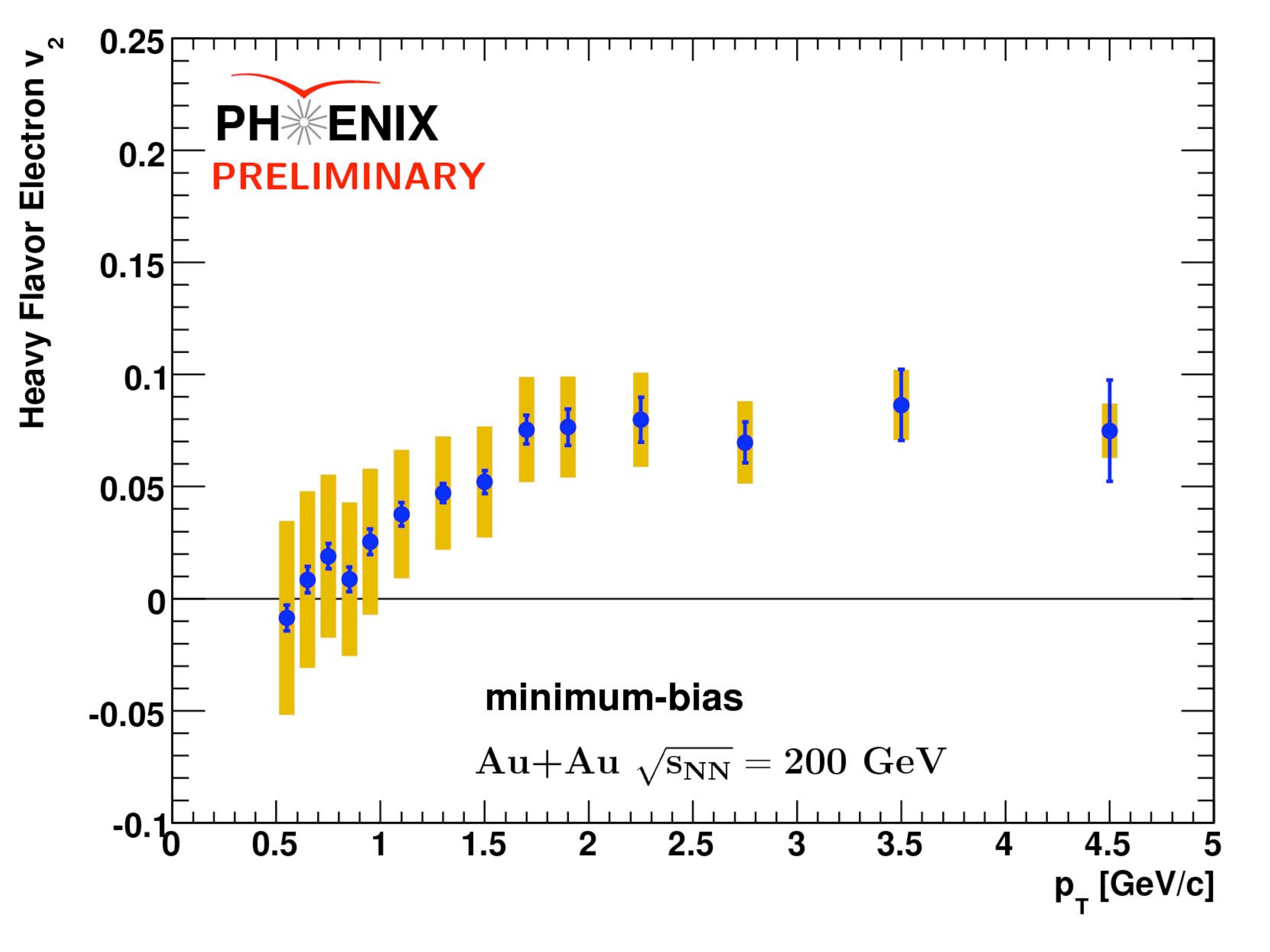} 
\end{tabular}
\end{center}\vspace*{-1.5pc}
\caption[]{a) (left) $R_{AA}$ (central Au+Au) b) (right) $v_2$ (minimum bias Au+Au) as a function of $p_T$ for direct-$e^{\pm}$ at $\sqrt{s_{NN}}=200$ GeV~\cite{PXcharmAA06}. }
\label{fig:fcrisis}\vspace*{-0.125in}
\end{figure}

The suppression of direct-single-$e^{\pm}$ in Au+Au collisions at RHIC (Fig.~\ref{fig:f7}) is even more dramatic as a function of $p_T\gsim 5$ GeV/c (Fig~\ref{fig:fcrisis}a) which indicates suppression of heavy quarks as large as that for $\pi^0$ (from light quarks and gluons) in the region where the $m\gsim 4$ GeV $b$-quarks dominate. Figure~\ref{fig:fcrisis}b  shows that heavy quarks exhibit collective flow ($v_2$), another indication of a very strong interaction with the medium. 
This observation calls into question the QCD radiative energy-loss explanation of jet-quenching~\cite{BSZARNS00}  because, naively, heavy quarks should radiate much less than light quarks and gluons in the medium. This issue has attracted much theoretical attention~\cite{egMGPL}. 

\pagebreak
\subsection{Zichichi to the rescue?}
  In September 2007, I read an article by Nino Zichichi, ``Yukawa's gold mine'', in the CERN Courier taken from his talk at the 2007 International Nuclear Physics meeting in Tokyo, Japan, in which he proposed:``We know that confinement produces masses of the order of a giga-electron-volt. Therefore, according to our present understanding, the QCD colourless condition cannot explain the heavy quark mass. However, since the origin of the quark masses is still not known, it cannot be excluded that in a QCD coloured world, the six quarks are all nearly massless and that the colourless condition is `flavour' dependent.'' 
  
  Nino's idea really excited me even though, or perhaps because, it appeared to overturn two of the major tenets of the Standard Model since it seemed to imply that: QCD isn't flavor blind;  the masses of quarks aren't given by the Higgs mechanism.  Massless $b$ and $c$ quarks in a color-charged medium would be the simplest way to explain the apparent equality of gluon, light quark and heavy quark suppression indicated by the equality of $R_{AA}$ for $\pi^0$ and direct single-$e^{\pm}$ in regions where both $c$ and $b$ quarks dominate. Furthermore RHIC and LHC-Ions are the only place in the Universe to test this idea. 
  
\begin{figure}[!h]
\hspace*{0.05\linewidth}\begin{minipage}{0.5\linewidth}
\includegraphics[width=\linewidth]{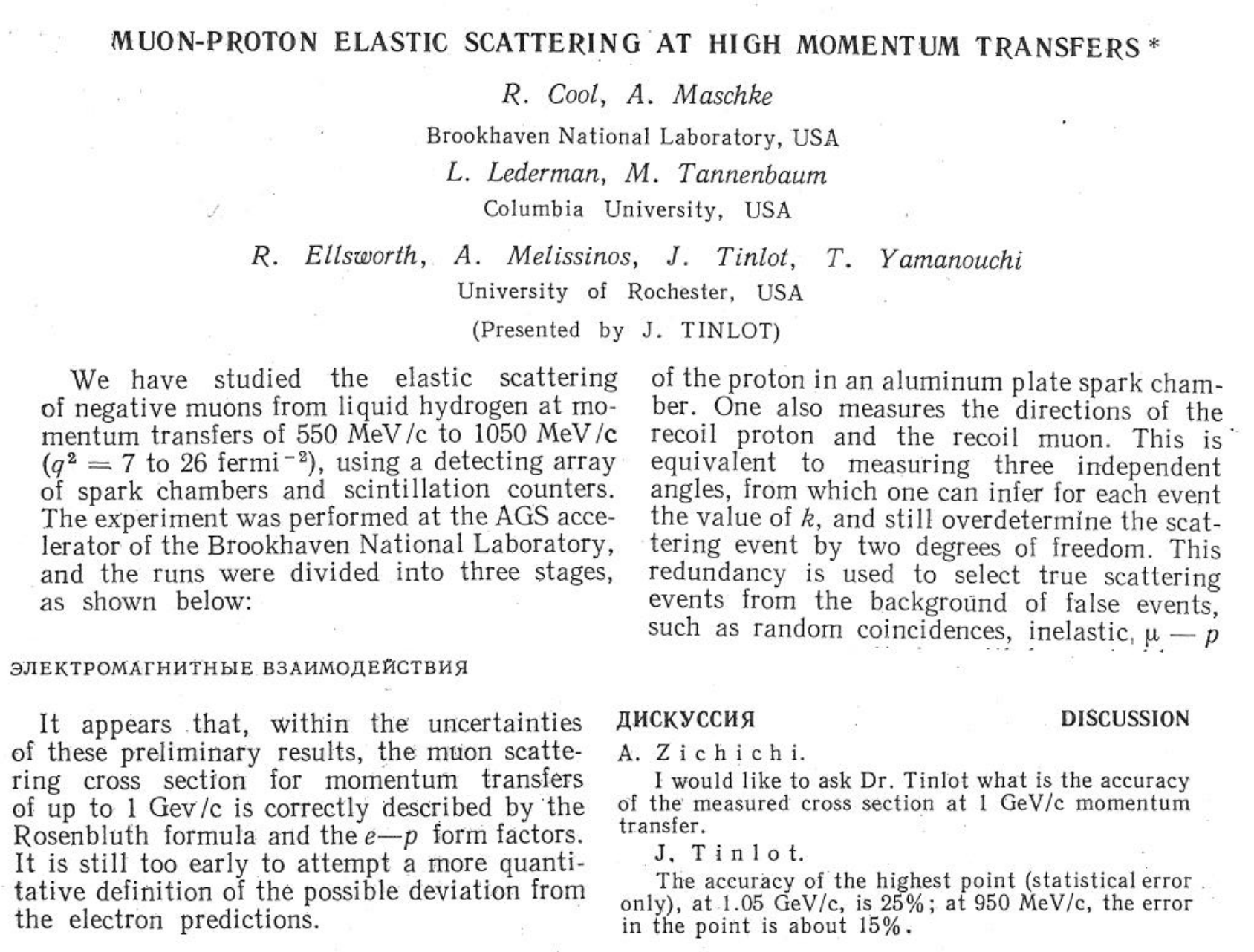}
\includegraphics[width=\linewidth]{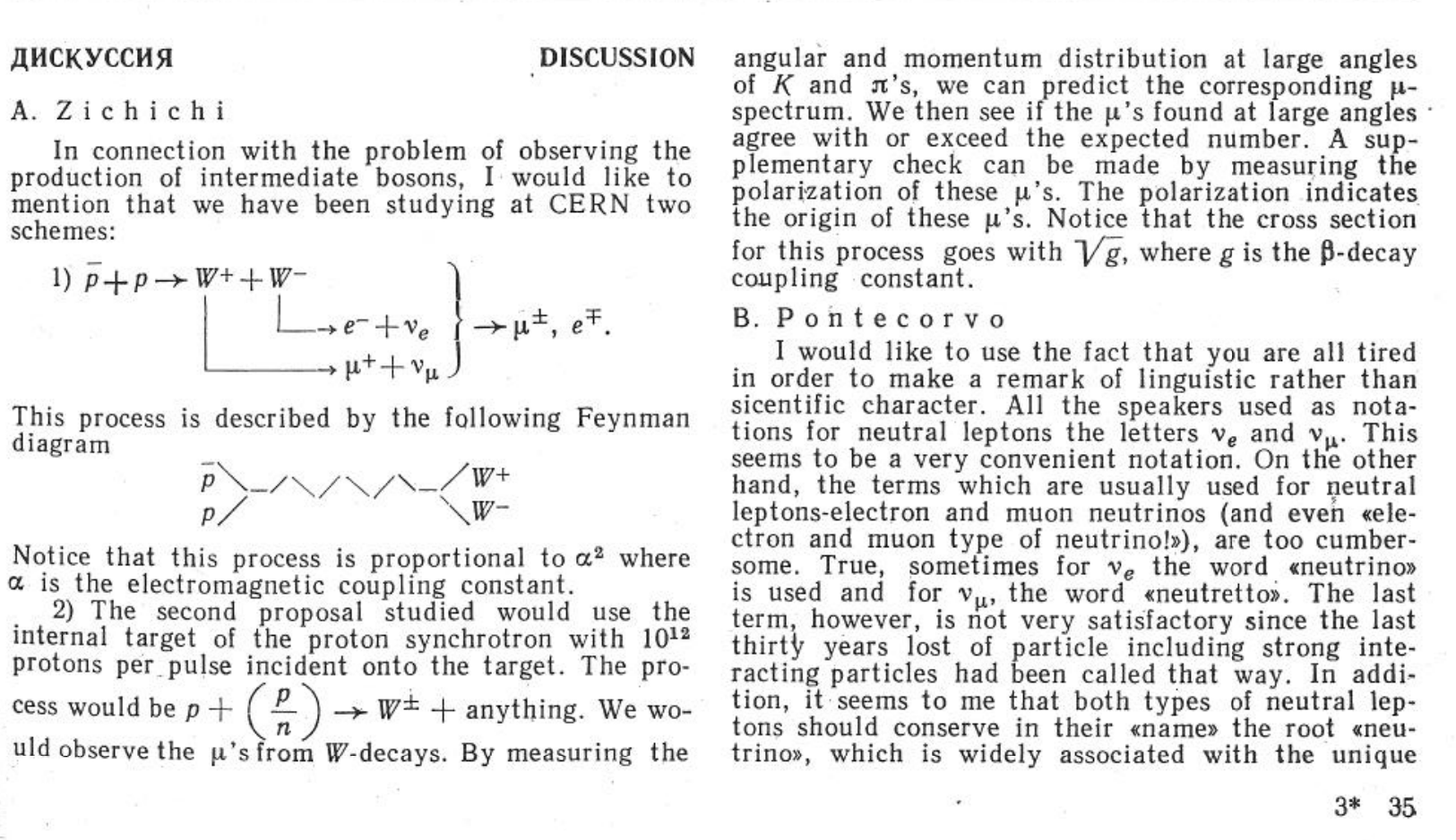}
\end{minipage}\hspace{1pc}%
\begin{minipage}{0.4\linewidth}
\includegraphics[width=\linewidth]{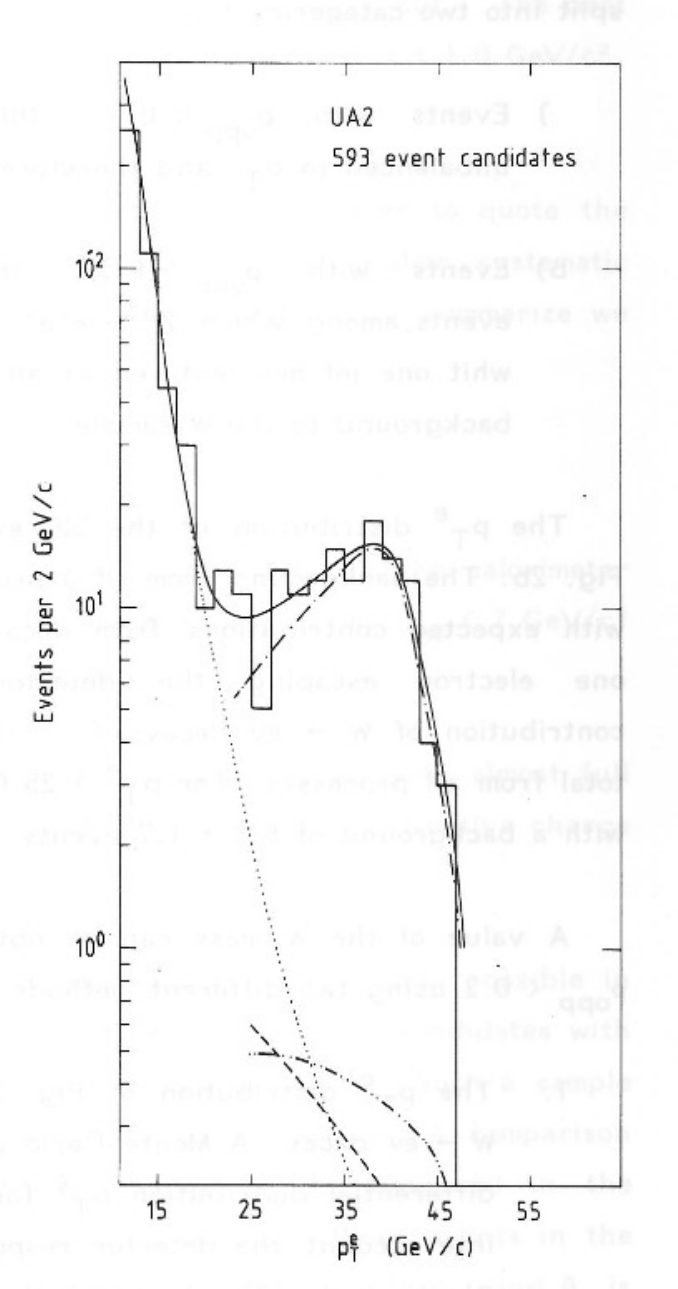}
\end{minipage} \vspace*{-1.0pc}
\caption[]{a)(left)Zichichi ICHEP 1964~\cite{MuPDubna64,NinoWDubna64} and b)(right) W~\cite{UA2WZ86} \label{fig:ZichichiW}}
\end{figure}

   It may seem surprising that I would be so quick to take Nino's idea so seriously. This confidence dates from my graduate student days when I checked the proceedings of the 12th ICHEP in Dubna, Russia in 1964 to see how my thesis results were reported~\cite{MuPDubna64} and I found several interesting questions and comments by an ``A. Zichichi'' printed in the proceedings (Fig.~\ref{fig:ZichichiW}). One comment about how to find the $W$ boson in p+p collisions~\cite{NinoWDubna64} deserves a verbatim quote because it was exactly how the $W$ was discovered at CERN 19 years later~\cite{Wdiscovery1,Wdiscovery2}: ``We would observe the $\mu$'s from W-decays. By measuring the angular and momentum distribution at large angles of K and $\pi$'s, we can predict the corresponding $\mu$-spectrum. We then see if the $\mu$'s found at large angles agree with or exceed the expected numbers.''

\begin{figure}[h]
\begin{center}
\includegraphics[width=0.95\linewidth]{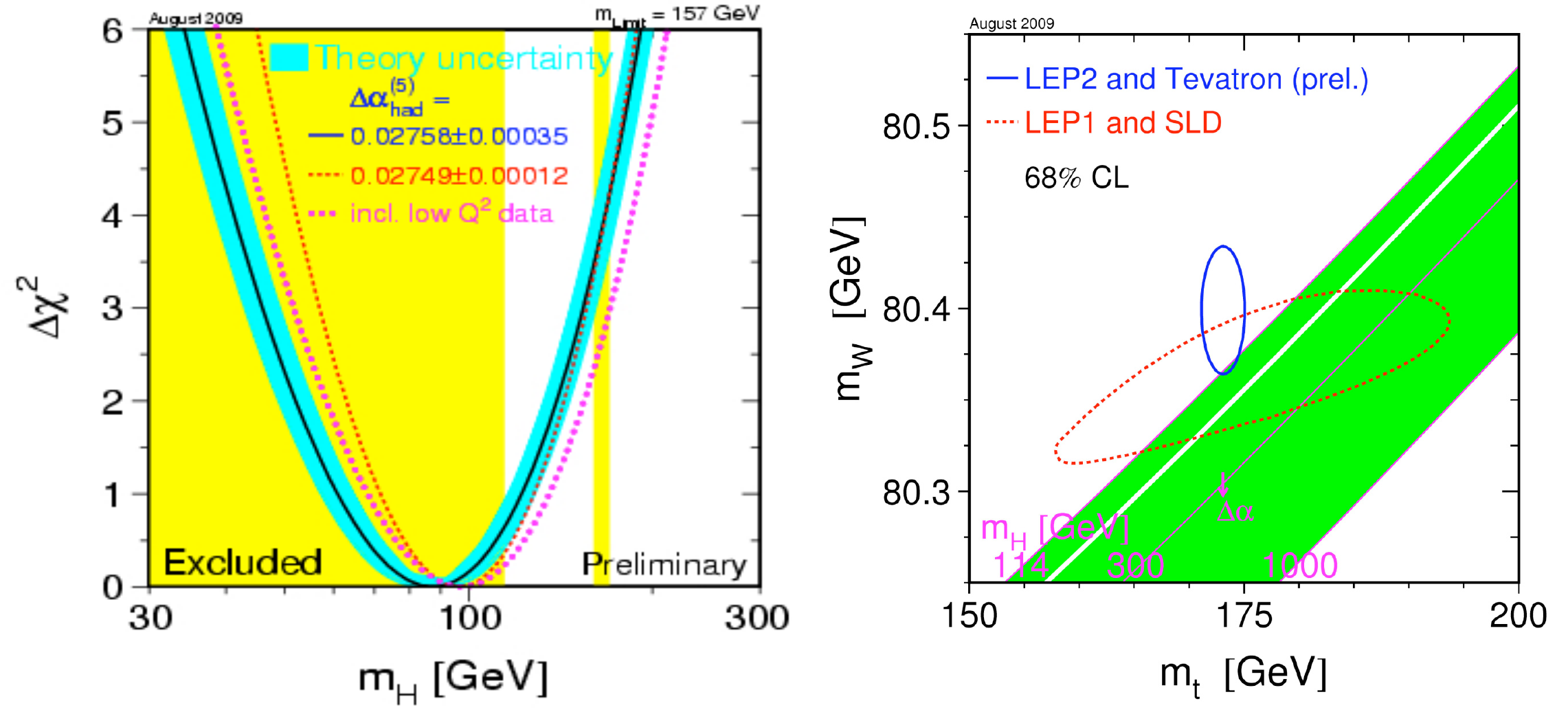}
\end{center}
\caption[]{Limits on Higgs mass from Tevatron and LEP, Summer 2009~\cite{TeVHiggs09}}
\label{fig:HiggsLimits}
\end{figure}

  Nino's idea seems much more reasonable to me than the string theory explanations of heavy-quark suppression (especially since they can't explain light-quark suppression). Nevertheless, just to be safe, I asked some distinguished theorists what they thought, with these results:
  \begin{itemize}
  \item Stan Brodsky:``Oh, you mean the Higgs field can't penetrate the QGP.''
 \item Rob Pisarski: ``You mean that the propagation of heavy and light quarks through the medium is the same.''
 \item Chris Quigg (Moriond 2008): ``The Higgs coupling to vector bosons $\gamma$, $W$, $Z$ is specified in the standard model and is a fundamental issue. One big question to be answered by the LHC is whether the Higgs gives mass to fermions or only to gauge bosons. The Yukawa couplings to fermions are put in by hand and are not required.'' ``What sets fermion masses, mixings?"
 \item Bill Marciano:``No change in the $t$-quark, $W$, Higgs mass relationship (Fig.~\ref{fig:HiggsLimits}) if there is no Yukawa coupling: but there could be other changes.''
 \end{itemize}
	 
	 Nino proposed to test his idea by shooting a proton beam through a QGP formed in a Pb+Pb collision at the LHC and seeing the proton `dissolved' by the QGP. My idea is to use the new PHENIX Silicon VTX detector, to be installed in 2010, to  map out, on an event-by-event basis, the di-hadron correlations from identified $b-\overline{b}$ di-jets, identified $c-\overline{c}$ di-jets, which do not originate from the vertex, and light quark and gluon di-jets, which originate from the vertex and can be measured with $\pi^0$-hadron correlations. A steepening of the slope of the $x_E$ distribution of heavy-quark correlations as in $\pi^0$-hadron correlations~\cite{ppg106} will confirm in detail (or falsify) whether the different flavors of quarks behave as if they have the same energy loss (hence mass) in a color-charged medium. If Nino's proposed effect is true, that the masses of fermions are not given by the Higgs particle, and we can confirm the effect at RHIC or LHC-Ions, this would be a case where we Relativistic Heavy Ion Physicists may have something unique to contribute at the most fundamental level to the Standard Model, which would constitute a ``transformational discovery.'' Of course the LHC could falsify this idea by finding the Higgs decay to $b-\bar{b}$ at the expected rate in p-p collisions. Clearly, there are exciting years ahead of us!

\ack {Research supported by U. S. Department of Energy, DE-AC02-98CH10886.}

\end{document}